%% file: paper_final.tex
\documentclass[%
nofootinbib,
 amsmath,amssymb,
 prd,
twocolumn
]{revtex4}
\usepackage{natbib}
\usepackage{url}
\usepackage{framed}
\usepackage{color}
\usepackage{hyperref}
\usepackage{graphicx}
\usepackage{dcolumn}
\usepackage{bm}
\usepackage{cleveref}
\usepackage{epsfig}
\usepackage[caption=false]{subfig}
\usepackage{times}
\usepackage{booktabs}

\newcommand{\hl}[1]{#1}{}

\newcommand{\e}[1]{\ensuremath{\times 10^{#1}}}
\newcommand{\dV}{\ensuremath{\textrm{dV}}}
\newcommand{\alfven}{Alfv\'{e}n }
\newcommand{\alfvenic}{Alfv\'{e}nic }

\renewcommand{\d}{\ensuremath{\textrm{d}}}
\usepackage{pgfplots}

\include{journals}
\newcolumntype{C}{>{$}c<{$}}
\AtBeginDocument{
\heavyrulewidth=.08em
\lightrulewidth=.05em
\cmidrulewidth=.03em
\belowrulesep=.65ex
\belowbottomsep=0pt
\aboverulesep=.4ex
\abovetopsep=0pt
\cmidrulesep=\doublerulesep
\cmidrulekern=.5em
\defaultaddspace=.5em
}
\begin{document}

\preprint{APS/123-QED}

\title{Nonhelical turbulence and the inverse transfer of energy: \\
A parameter study}%
\author{Johannes Reppin}
 \email{johannes.reppin@hs.uni-hamburg.de}
\author{Robi Banerjee}%
 \email{rbanerjee@hs.uni-hamburg.de}
\affiliation{%
 Hamburger Sternwarte\\
 Gojenbergsweg 112, 21029 Hamburg, Germany
}%

\date{\today}

\begin{abstract}

    We explore the phenomenon of the \hl{recently discovered {\em inverse transfer} of energy
  from small to large scales in decaying magnetohydrodynamical
    turbulence by Brandenburg et al. (2015) even for nonhelical magnetic fields}.  
    For this investigation we mainly employ the \textit{Pencil-Code} performing a
  parameter study, where we vary the Prandtl number, the kinematic
  viscosity and the initial spectrum.
  We find that in order to get a decay which exhibits this
  \textit{inverse transfer}, large Reynolds numbers
  ($\mathcal{O}\sim 10^{3}$) are needed and low Prandtl numbers of the
  order unity $Pr=1$ are preferred.
  Compared to \textit{helical} MHD turbulence, though, the
  \textit{inverse transfer} is much less efficient in transferring
  magnetic energy to larger scales than the well-known effect of the \textit{inverse cascade}.
  Hence, applying the \textit{inverse transfer} to the magnetic field evolution in
  the Early Universe, we question whether the \textit{nonhelical} inverse
  transfer is effective enough to explain the observed void magnetic
  fields if a magneto-genesis scenario during the electroweak phase
  transition is assumed.
\end{abstract}

\pacs{Valid PACS appear here}
\maketitle

\section{Introduction}

Magnetohydrodynamic turbulence offers a rich variety of physical
phenomena and hence is still a field of intense research.  In
comparison to pure (incompressible) hydrodynamical turbulence, the
presence of magnetic fields introduces additional complexity to the
problem, changing the picture from the classical turbulence theory
introduced by Kolmogorov \citep{Kolmogorov41}.

Hence, MHD turbulence has long been an area of interest.
For example, a large scale background field or the field on the
largest eddy-containing scales could give rise to modifications of the
small-scale fluctuations compared to pure hydrodynamic turbulence.
Iroshnikov \cite{Iroshnikov1964} \& Kraichnan
\cite{Kraichnan1965b} firstly introduced a modified theoretical
description where only waves of opposite directions interact.
This interaction is then governed by the \alfven timescale
$\tau_A\sim l/v_A$ which is shorter than the eddy-distortion time
$\tau_l$ considered otherwise.  This introduces an additional factor
of $\tau_l/\tau_A$ in the energy-transfer time which is used in the
derivation of the hydrodynamic turbulence theory.  The \alfven effect
\citep{Iroshnikov1964,Kraichnan1965b} causes the inertial-range
scaling to differ from classical HD turbulence, effectively leading to
a more shallow spectrum of $E\sim k^{-3/2}$, rather than $E\sim k^{-5/3}$.

The theory of \alfvenic wave interaction was extended by Sridhar \&
Goldreich and Goldreich \& Sridhar 
\citep{SridharGoldreich94, SridharGoldreich95} to include interactions
of multiple \alfvenic wave modes.  It then follows that 3-mode wave
interactions do not give rise to resonances which then leads to the
conclusion of a failure of the IK theory. \hl{A first complete discussion of MHD
turbulence with resonant interactions was discussed by Galtier et. al (2000)} \cite{Galtier2000}.

The detailed analysis of MHD turbulence including resonant 4-wave
interactions gives rise to a steeper spectrum of
$E_k\propto k_{\|}^{-2}$, but also a highly anisotropic spectrum where
one has to differentiate between the perpendicular and the parallel
parts of the energy spectrum.  Cho \& Vishniac
\cite{ChoVishniac2000} and more recently also Beresnyak
\cite{Beresnyak15} conclude in their analysis that a theoretically
derived Sridhar \& Goldreich spectrum agrees with their numerical
simulations.  See the book by Biskamp \cite{BiskampBook} for a
a more detailed discussion about the effects of \alfvenic waves in
turbulent fields.

Another important phenomenon is observed when the field exhibits
\textit{magnetic helicity}. In this case it is well known that the
decay is drastically different where one observes an increase of
magnetic energy on large scales and hence, a dynamical growth of the
correlation length (see e.g. Pouquet et al., Christensson et
al., \cite{Pouquet1976,Christensson01}). This effect of an
\textit{inverse cascade} is due to the well conserved helicity during
the turbulent decay (see also \cref{sec:hel_decay} in this work).
  
Without the presence of helicity, earlier studies by Batchelor \citep{Batchelor50},
Saffman \citep{Saffman67},  Banerjee \& Jedamzik \citep{Banerjee04} and Sethi et al. \citep{Sethi05} showed that for blue magnetic field spectra, \hl{i.e. a spectrum which rises for large Fourier modes}, the coherence length also increases but this happens 
by the damping of small-scale fluctuations, leaving only large-scale fluctuations.
In this case, the decay law for the magnetic energy and the growth rate of the coherence length
depends on the large scale spectral index of the magnetic field
fluctuations. 
Also, other numerical studies confirmed that
the peak of the magnetic spectrum moves along the large scale spectrum
while the small-scale fluctuations decay
(e.g. \cite{Campanelli07,Saveliev12,Saveliev13}). 

Recently, Brandenburg et al. \cite{Brandenburg15} suggested
that, even without helicity, the magnetic energy can increase on scales
larger than the initial integral scale, and the coherence length can
moderately grow through an effect similar to the helical case. 
But this nonhelical \textit{inverse transfer} requires high Reynolds
numbers. Hence, previous studies of nonhelical MHD turbulence decay
have not seen this effect clearly \citep{MuellerBiskamp99,
  Christensson01, Banerjee04b, Kalelkar04} whereas latest numerical studies seem
to confirm the result by Brandenburg et al.
\cite{Zrake2014,Linkmann16}. 

If the effect of the inverse transfer is proven to be universal for
large Reynolds number regimes, it would have a large impact for
magnetic field evolution, in particular, during the Early
Universe. Here, slight changes of the turbulent decay law (which is
typically a power law of the time $t$) will result in different field
strengths and coherence lengths at later epochs. For instance,
assuming a magneto-genesis scenario during the electroweak (EW) phase
transition, Wagstaff et al. \cite{Wagstaff16} \hl{showed that the
decay of the magnetic field will be too fast to explain the weak lower bounds
of the magnetic fields in the voids of galaxies as inferred from Fermi
observations of TeV Blazars} \cite{Neronov10, Taylor11}. Without the
effect of the inverse transfer the magnetic energy will decay as
$E_B \propto t^{-10/7}$ (see e.g. \cite{Banerjee04b, Sethi05}) if a
causal magnetic field spectrum with no helicity is assumed
\cite{Durrer03}. Otherwise, according to the decay law by the effect
of the inverse transfer, the magnetic energy decays as $E_B \propto
t^{-1}$, leaving strong enough present-day magnetic fields to explain
the fields in voids of galaxies \cite{Kahniashvili13}. 

Motivated by the work by Brandenburg et. al (2015) \cite{Brandenburg15},
we performed a detailed numerical investigation to test the regimes where one can
expect an efficient inverse transfer of energy to larger scales during
the decay of magnetic fields.
Unlike in the case of helical magnetic fields, where helicity is a
conserved quantity and energy is transferred to larger scales by an
inverse cascade, the physical reasoning of the non-helical
\textit{inverse transfer} is not understood (see appendix of this work
and supplement of \cite{Brandenburg15}). Our goal is to shed some light on
this phenomenon. For this reason, we mainly use the well established
Pencil Code \cite{PENCIL10}, which was also used in the original study
by Brandenburg et al. \cite{Brandenburg15}, where we vary the
Reynolds number, the Prandtl number as well as the initial spectra of
the stochastic magnetic field.

This work starts off with a discussion of helical and nonhelical
turbulence in \cref{sec:mhdturb}.  In \cref{sec:nummethods} we
describe the details of our numerical setup, the analysis methods and
the run-time parameters. We then present the simulation results in
\cref{sec:results} where we discuss the impact of the variation of
the viscosity parameter and the Prandtl number as well as the
influence of the initial spectrum on the \textit{inverse transfer}.
Furthermore, we present a run with the \textit{Zeus-MP2} Code, where
we don't see enough evidence for the \textit{inverse transfer} and
discuss how that could be linked to the numerical integration
scheme. We conclude in \cref{sec:conclusions} with the discussion
of our main findings and their impact on causally generated fields in
the early Universe questioning the effect of the inverse transfer due
to large expected Prandtl numbers. 
\section{MHD Turbulence}
\label{sec:mhdturb}
In this section we briefly summarize the general properties of turbulence, with a focus on decaying magnetohydrodynamic turbulence. 
A detailed description of the numerical implementation follows in \cref{sec:hypervisc}. 
\begin{figure*}%
    \centering
    \subfloat[Slices of the nonhelical run, $\mathcal{H}_0=0$. The size of the eddies grow through the decay of small-scale fluctuations as well as and by the effect of the \textit{inverse transfer}]
    {\begin{minipage}[t]{0.4\linewidth}
    \includegraphics[width=\textwidth]{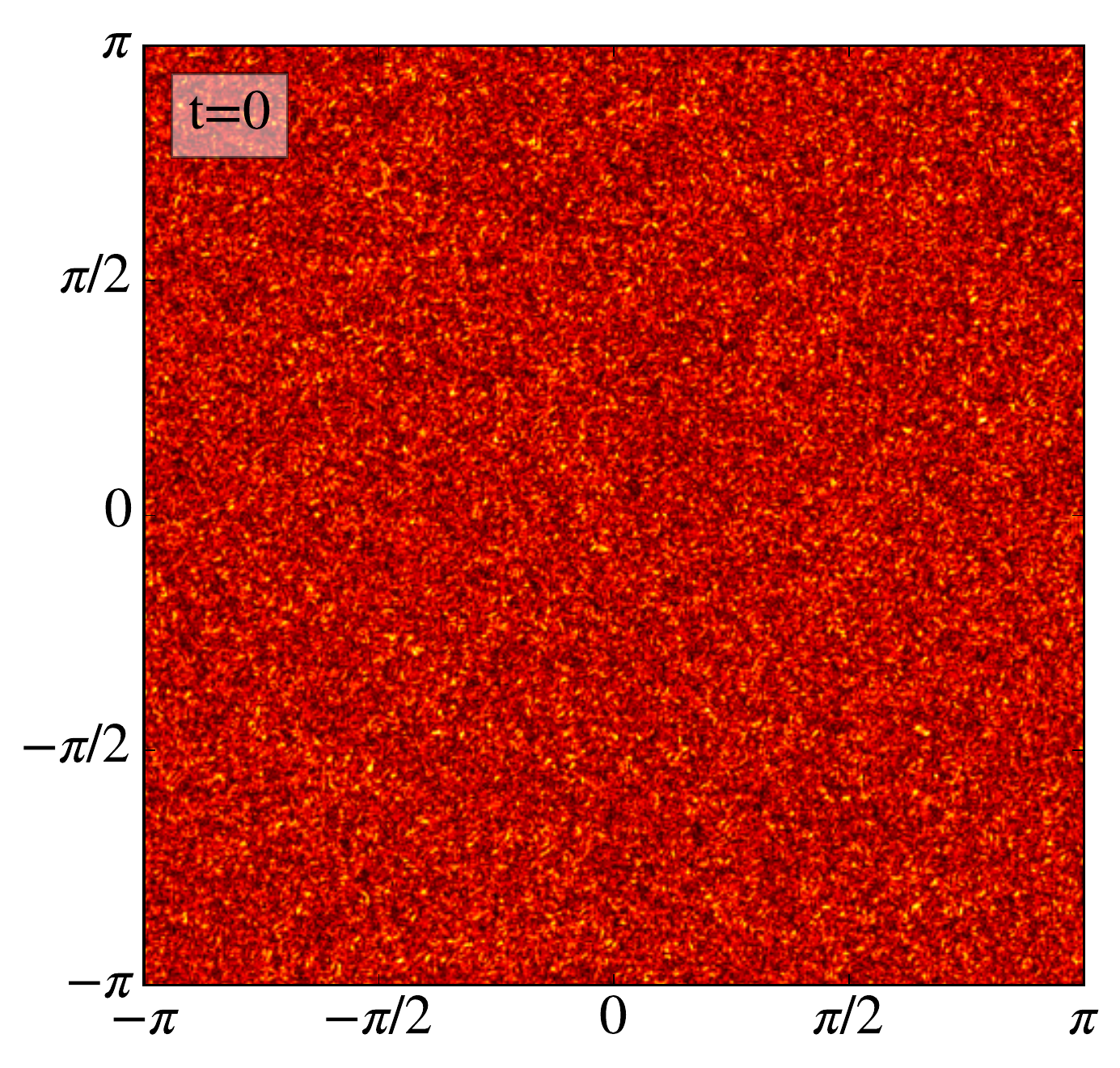}
    \end{minipage}
    \hfill
    \begin{minipage}[t]{0.4\linewidth}
    \includegraphics[width=\textwidth]{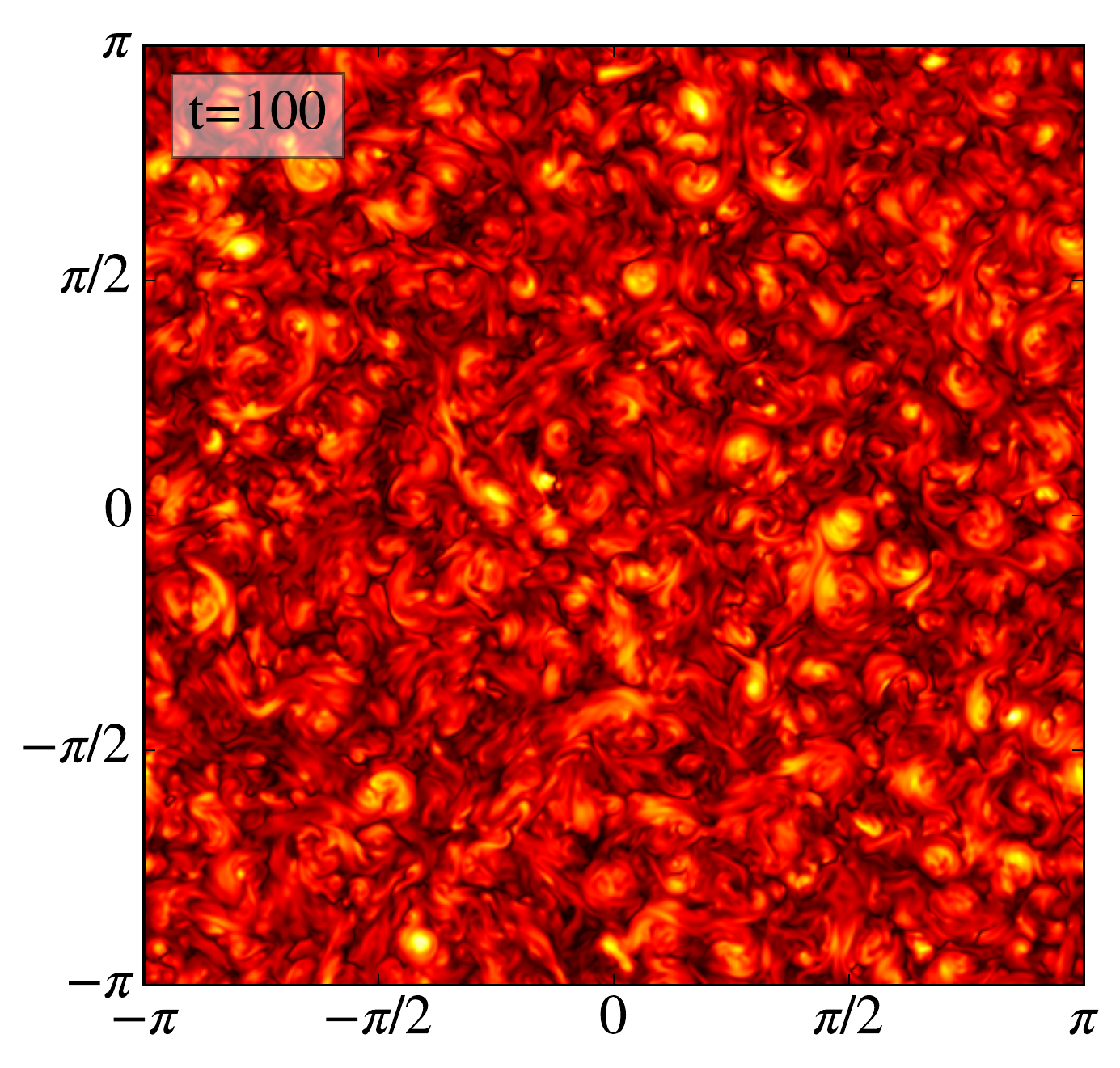}
    \end{minipage} \label{fig:slices_nonhel}} \\
    \subfloat[Slices of the maximally helical run. The eddies grow through an \textit{inverse cascade} of magnetic energy. The eddies are now about $1/5$ of the box size]
    {\begin{minipage}[t]{0.4\linewidth}
    \includegraphics[width=\textwidth]{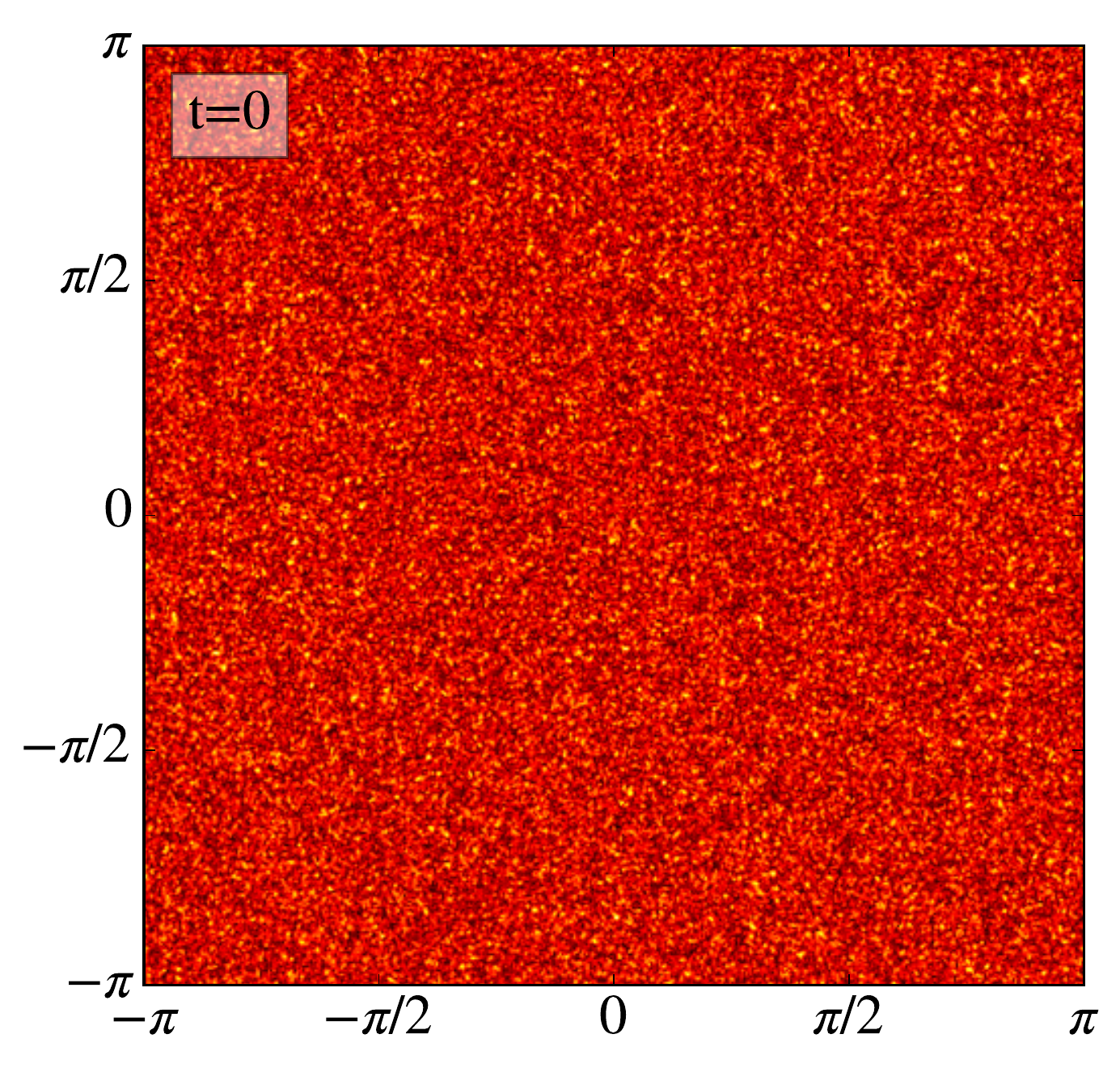}
    \end{minipage}
    \hfill
    \begin{minipage}[t]{0.4\linewidth}
    \includegraphics[width=\textwidth]{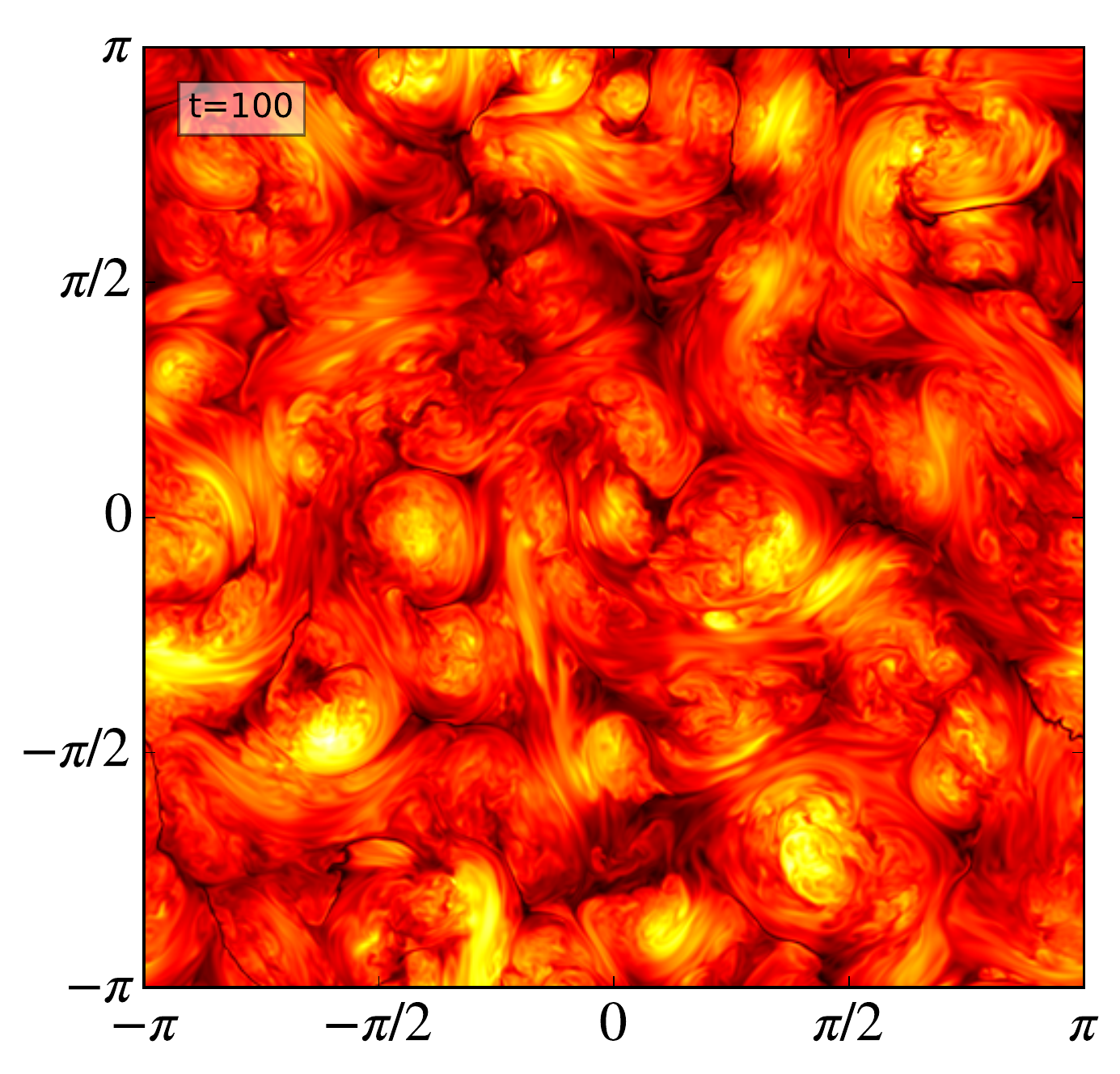}
    \end{minipage}\label{fig:slices_hel}}
    \caption{Slices of the \textit{xy-plane} of the magnetic field strength $|B|=\sqrt{B_x^2 + B_y^2 + B_z^2}$ at the initial time and at $t=100$. The color scale has been adjusted for each panel, the energy has decayed significantly at the later snapshot.}%
    \label{fig:turb_slices}%
\end{figure*}
\subsection{Nonhelical Decay}
The general picture of MHD turbulence is similar to the case of pure hydrodynamic turbulence. 
The magnetic field decays by excitation of velocity fluctuations which in turn decay through a turbulent cascade and ultimately by dissipation into heat.
The decay rate depends on the initial magnetic power spectrum $E_k$, which is defined by:
\begin{align}
E &= \frac{1}{2}\int \textbf{B(\textbf{k})}^2 \textrm{d}^3k \nonumber\\ 
  &= \int  k^2 |B(k)|^2\d\Omega \d k \nonumber\\
  &= \int \d k E_k = \int k E_k \d \ln k
\label{eq:powerspectrum}
\end{align}
where $E$ is the total magnetic field energy.
Often it is assumed that the power spectrum, i.e. the magnetic energy per wavenumber bin, is reasonably isotropic and given by a power law: 
\begin{equation}
E_k\propto k^{n},
\end{equation}
where $n$ is the spectral index at large scales, i.e. $k<k_I$ and $L_I=2\pi/k_I$ is the integral scale of the magnetic field, the scale at which $E_k$ has its maximum.
Note that here we use the integral scale, the coherence length and the correlation length interchangeably.
If we assume most of the energy is located at the integral scale, the total energy can be estimated as  
\begin{equation}
E=\int k E_k\d \ln k \approx k_I E_I,
\end{equation} 
where $E_I$ denotes the spectral energy at $k_I$.

During a Kolmogorov cascade, the largest eddies have the longest relaxation time, also called eddy turnover time $\tau_k \sim l_k/v_k$, which is why the integral scale dominates the rate at which the decay occurs. 
One can then derive a decay law for the magnetic energy: 
\begin{equation}
\label{eq:energydecay}
E(t) = E_0(1+t/\tau_0)^{-\frac{2(n+1)}{3+n}},
\end{equation}
where $\tau_0$ is the initial eddy turnover time \citep[e.g.][]{Banerjee04b, Sethi05}.

Note that our index $n$ is different from \citep{Banerjee04b}, who used $E_k\sim k^{n'},\ n=n'-1$. 

In principal $n$ can take any value.
The simplest case would come from an average of randomly distributed magnetic dipoles, which results in a spectral index of $n=2$  \citep[e.g.][]{Hogan83}. This will result in a decay law of $E\propto t^{-6/5}$, which is also known as Saffman's law \citep{Saffman67}.
Another commonly assumed value for a blue spectrum is $n=4$, which represents a causally generated magnetic field during a phase transition in the early Universe \citep{Durrer03}, which leads to a decay law of $E\sim t^{-10/7}$. 
Furthermore, a weak magnetic field which gets amplified via the small-scale turbulent dynamo will develop a Kazantsev slope ($n$ = 3/2) at large scales \citep{Kazantsev68}.
\subsection{Helical Decay} \label{sec:hel_decay}
\begin{figure}
    \input{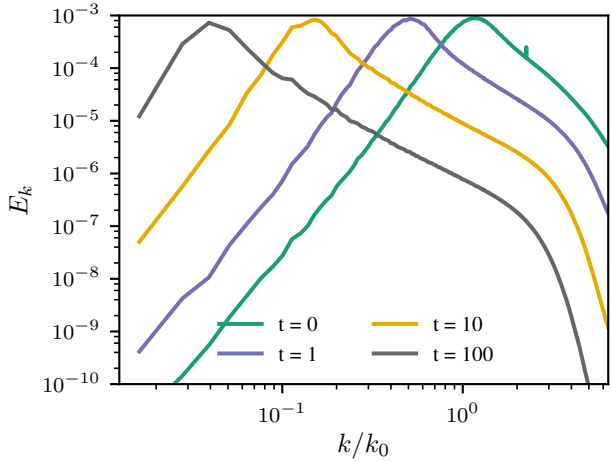}
    \caption{Magnetic power spectra of a run with maximal helicity. While the magnetic energy decays, the peak of the power spectrum does not, but shifts to lower $k$. This \textit{inverse cascade} only occurs for maximally helical fields, where $H_k\sim E_k$.}
\label{fig:helical_mag_spec}
\end{figure}
A magnetic field can also exhibit helicity, which is a measure of the twisting of the magnetic field lines. 
The helicity is defined as the volume integral
\begin{equation}
\mathcal{H} = \int \mathbf{A} \cdot \mathbf{B}\ \dV,
\label{eq:helicity}
\end{equation}
 where $\mathbf{A}$ is the vector potential of the magnetic field. 

In a highly conducting medium the helicity is nearly conserved, i.e. it is much better conserved than the magnetic energy. 
$\mathcal{H}$ has the dimension of magnetic energy times the correlation length, i.e. $\mathcal{H}\sim B^2L \sim E_BL$. Hence, $\mathcal{H}$ has the same dimension as the magnetic power spectrum $E_k$. This means that with decaying magnetic energy for a field which is maximally helical, i.e. $\mathcal{H}_{max}\approx E_k$, the correlation length has to increase to ensure helicity conservation. 
The result is a transfer of magnetic energy from smaller to larger scales, i.e. the field evolves via an \textit{inverse cascade}.
This has also been know from numerical studies \citep[e.g.][]{MuellerBiskamp99, Christensson01, Banerjee04b, Mueller12}.
Here, we also illustrate the evolution of a maximally helical field
in \cref{fig:helical_mag_spec}, where we show the magnetic power spectra at different to compare the nonhelical runs performed for this study. As expected, the peak of the power spectrum remains nearly constant.
\section{Numerical Methods}
\label{sec:nummethods}
For our study we employ the well-established \textsc{Pencil-Code}\footnote{http://pencil-code.nordita.org} which solves the compressive MHD equations with an isothermal equation of state. 
The set of relevant MHD equations are summarized as follows: 
\begin{align}
\frac{\textrm{D}\ln\rho}{\textrm{D} t} &= -\nabla \cdot \mathbf{u} \label{eq:MHD1} \\
\frac{\textrm{D}\mathbf{u}}{\textrm{D}t} &= -\rho^{-1}\nabla p + \frac{\mathbf{j}\times \mathbf{B}}{\rho} + \mathbf{f}_\textrm{visc}  \\
\frac{\partial\mathbf{A}}{\partial t} &=  \mathbf{u}\times \mathbf{B} - \eta\ \mathbf{j} \label{eq:MHD3}
\end{align}
Where $\textrm{D}/\textrm{D}t = \partial/\partial t + \mathbf{u} \cdot \nabla \label{eq:convectderiv}$ is the convective derivative and $\mathbf{f}_\textrm{visc}$ is a viscous force.
The code uses a sixth-order finite differences scheme which uses the logarithmic density $\ln \rho$, the velocity $\mathbf{u}$ and the vector potential $\mathbf{A}$ as primitive variables. 
It advances the magnetic vector potential, where the magnetic field is $\textbf{B}=\nabla \times\textbf{A}$ and  $\mathbf{j}=\nabla \times \mathbf{B}$ is the MHD current.
This results in the divergence free condition $\nabla\cdot \textbf{B}=0$ being inherently fulfilled.
\subsection{Hyperviscosity}
\label{sec:hypervisc}
A numerical technique we use in a subset of the simulations is the hyperviscosity  \citep{Haugen04}.
It has the form of a high-order derivative of the velocity field:
\begin{equation}
\label{eq:hyper3}
\textbf{f}_{\textrm{hyper} 3} = \nu_3 \nabla^6 \textbf{u}
\end{equation}
It is a replacement of the standard Laplacian viscosity term which appears in the Navier-Stokes equation:
\begin{equation}
\label{eq:laplacian}
\textbf{f}_{\textrm{visc}}=\nu \nabla^2 \textbf{u}
\end{equation}
It has been shown that by using the hyperviscosity instead of Laplacian viscosity, the inertial range of the simulation can be largely increased \citep{Haugen04}.
This is also reflected by an increase of the Reynolds number without increasing the numerical resolution of the simulation. 
\subsection{Reynolds Numbers in the Simulations}
The study by Brandenburg et al. indicated that the \textit{inverse transfer} is only observable for large Reynolds numbers, i.e. $Re > 10^3$.
In the case of the Laplacian type viscous force, the Reynolds number is defined as follows:
\begin{equation}
\label{eq:reynolds}
Re = \frac{v\cdot L_I}{\nu} = \frac{v\cdot 2\pi}{k_I\nu} 
\end{equation}
Here, we use $B_{rms}$ as an estimate for the velocity fluctuations $v$, the integral scale $L_I$ and the kinetic viscosity $\nu$.
We vary the parameter $\nu$ in the range $\nu=1\e{-4}\dots 5\e{-6}$.
This range results in Reynolds numbers from roughly $100$ to  $2\times10^3$.

In the case of hyperviscosity, the Reynolds number at the Nyquist frequency can be adjusted to be $5$ to $7$ \citep{Haugen04}.
In practice, this results in an effectively much larger Reynolds number for the simulation compared to the Laplacian case.

If one assumes Kolmogorov type turbulence in numerical simulations, one can get an estimate for the Reynolds number from the dissipation scale, given by the Nyquist wavenumber $k_{Ny}$, and the integral scale \citep{Haugen04a}.
\begin{equation}
Re = \left(\frac{k_{Ny}}{k_I}\right)^{4/3}
\end{equation}
This gives us values of $Re = 2\e{2}$ if we use $k_{Ny}=512$ and an integral scale of $k_I\approx 10$.
Using this estimate underestimates the Reynolds number, but it is independent of the implementation of the viscosity in our simulation which is useful as a second indicator of the expected turbulence. 
A more exact Reynolds number cannot be given, because the Reynolds number depends strongly on the numerical methods like the implementation of the viscosity. 
Nevertheless, high $Re$ are needed to properly resolve the effect of the \textit{inverse transfer}, as we will discuss in  \cref{sec:results}.
\subsection{Initial Conditions}
We generate our initial conditions in Fourier space to set up a specific magnetic power spectrum.
Our spectra have two parts, the part on large scales is a blue spectrum, the part on smaller scales is a decreasing spectrum, the division being at the integral scale, i.e. $L_I = 1/k_I$.
We also vary this blue part of the spectrum, in particular 
\begin{alignat}{2}
E_k &\sim k^n\  &&\textrm{for}\ k\leq k_I  \\
E_k &\sim k^{-5/3}\  &&\textrm{for}\ k>k_I. \nonumber
\end{alignat}
Where we assume a Kolmogorov spectrum on small scales and $n=4$ for most of the cases.
We set $k_I=80$ in order to achieve a reasonable separation of scales. 

Initially, we set the velocity field to zero to ensure a fully magnetically driven turbulence. 
The initial \textit{rms} magnetic field strength is set to $B_{0}=0.3$, corresponding to an energy of $E_B=4.5\times 10^{-2}$.
The sound speed is $c_s=1$ and the density is set to $\rho_0=1$. 
\hl{We plot our time series normalized to the initial \alfven time $\tau_0=(v_{A,0}\cdot k_{0})^{-1}$, with the initial \alfven velocity $v_{A,0}$ and the integral scale of the initial conditions $k_0=k_{I,t=0}$. }
Also, the magnetic energy spectra are normalized to the initial integral scale $k_0$.

Apart from one comparison run with $\mathcal{H}= \mathcal{H}_\textrm{max}$ (see \cref{fig:helical_mag_spec}) we initialize the magnetic field with zero helicity.

Furthermore, we also ran one simulation where we initiated the magnetic field at a single scale (at $k=300$) so that a \textit{natural} magnetic field spectrum was established (see e.g. Saveliev et al. \citep{Saveliev12}).
Such a delta peak spectrum can be related, for example, to a phase transition scenario that works on a specific scale $k_\delta$. 
\subsection{Correlation Length}
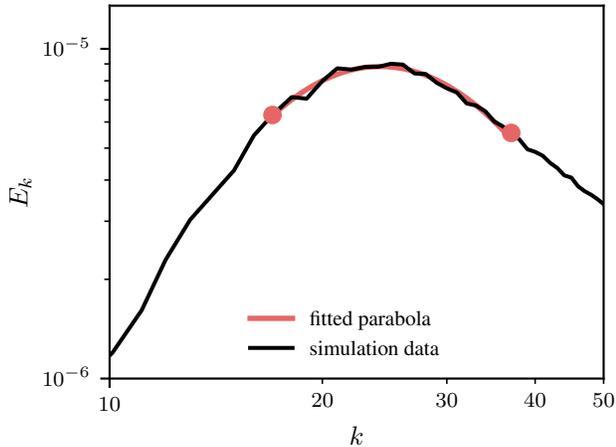
\begin{figure}
    \centering
    \input{fitting_example_nonhel_k80.pgf}
    \caption{Fitting of a parabola in log-space to determine the value for the wavenumber $k_I$ which defines the correlation length $L_I=1/k_I$, i.e. the integral scale.}
    \label{fig:example_fitting}
\end{figure}
We use a fitting function to the one-dimensional power spectrum to calculate the time evolution of the integral scale.
We fit a parabola in log space around the peak values of the power spectrum with a roughly equal interval in k-space.  
The peak of our parabola defines the correlation length of the magnetic field (see \cref{fig:example_fitting}).
\hl{This is equivalent to the analytic expression of $k_I^{-1}=\int k^{-1}E_M(k,t)/E(t)dk$, our spectral fitting approach being more descriptive, though.}
We discuss the temporal evolution of this scale $k_I$, or rather its inverse, the length scale $L_{\textrm{corr}}=1/k_I$  in \cref{sec:results}. 
For \textit{helical} MHD turbulence this value decreases over time while at the same time keeping a constant value of its peak energy $E(k_I)$. 
Previous work on nonhelical turbulence has shown that this is not the same when negligible helicity is present in the field's configuration.
\section{Results}
\label{sec:results}

\begin{table}
    \centering
    \begin{tabular}{ccc}
        \toprule
	{\bfseries Run }    & {\bfseries Parameters}  & {\bfseries Inverse Transfer} \\
	    \midrule
	{\bfseries viscosity} & $\nu$  &  \\
		Visc1  & $1\e{-4}$  &   decay only \\
		Visc2  & $5\e{-5}$  &   weak  \\
		Visc3  & $1\e{-5}$  &   medium \\
		Visc4  & $5\e{-6}$  & strong \\
		Hyper1 & $5\e{-15}$ &    strong \\
		Hyper2 & $2\e{-14}$ &    strong \\
	    \midrule
	{\bfseries Prandtl number} & $Pr_M$  &  \\
		Prandtl1  & 1000  &  weak \\
		Prandtl2  & 100   & medium \\
		Prandtl3  & 10    & strong \\
		Prandtl4  & 1     & strong \\
	    \midrule
	{\bfseries initial slope} & index $n$ &  \\
		Slope1 & $E_k\sim k^{1/2}$  & decay only \\ 
		Slope2 & $E_k\sim k^{1}$    & decay only \\ 
		Slope3 & $E_k\sim k^{2}$    & decay only \\ 
		Slope4 & $E_k\sim k^{3}$    & weak       \\ 
		Slope5 & $E_k\sim k^{4}$    & strong     \\ 
		Slope6 & $E_k\sim k^{6}$    & $E_k\rightarrow k^4 \rightarrow$ strong\\  
    \bottomrule
\end{tabular}
   \caption{Overview of the simulations performed and when the inverse transfer of energy occurred at which parameter set. The right-most column describes if the run exhibits the inverse transfer effect or if the parameters do not allow it. Note the last row where the spectrum first decays to a \textit{causal} form $E\sim k^4$ and then decays in the same way, with a strong inverse transfer.}
   \label{tab:parameters}
\end{table}
In this section we present the results from our 3D simulations. 
To get a general impression on the field evolution, we show a comparison of a helical and a non-helical run in \cref{fig:turb_slices}.
In \cref{tab:parameters} we give an overview of our simulations and their main results.
The simulations are carried out with a resolution of $1024^3$ if not otherwise stated. 
\subsection{Varying the Viscosity Parameter}
\label{sec:visc_parameter}
\begin{figure*}
    \centering
    \input{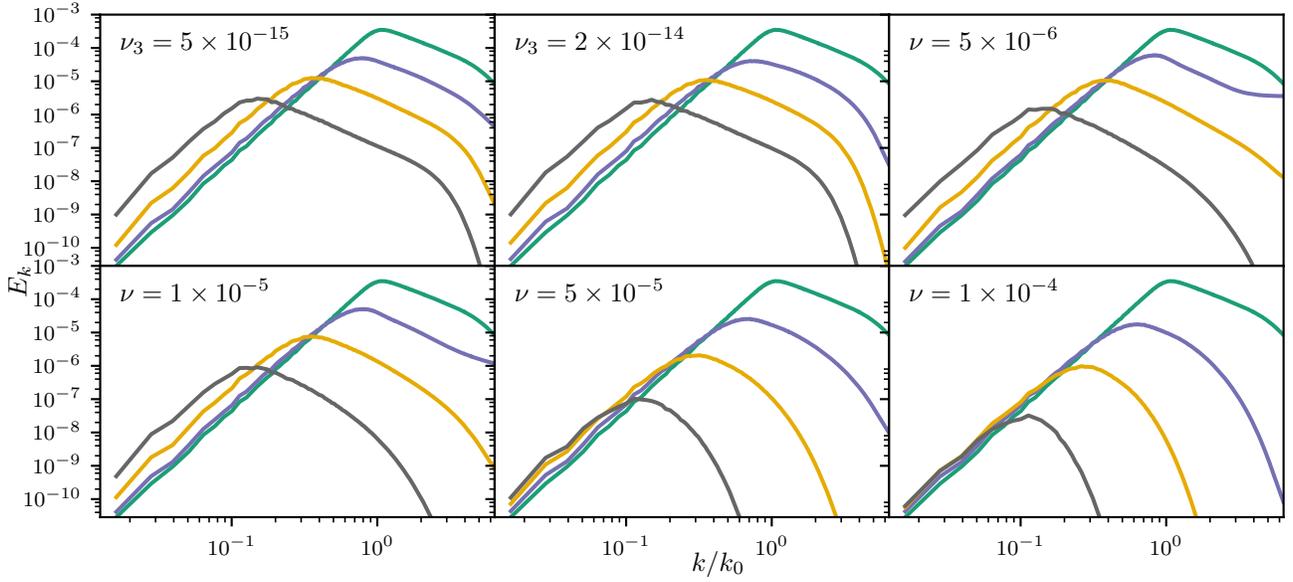}
    \caption{Magnetic power spectra for runs with different viscosities. 
    The hyperviscosity runs are shown in the two upper left panels. High Reynolds numbers, i.e. low viscosities, are needed to observe the effect of non-helical \textit{inverse transfer} which dynamically increases the energy on large scales. \hl{Spectra are shown for simulation times $t=0,20,2\e{2}\ \&\ 2\e{3}\ \tau_0$, where the green (uppermost) spectrum is the initial condition $t=0$. All runs have the Prandlt number $\textrm{Pr}=1$.}} 
    \label{fig:viscosity_comparison}
\end{figure*}
\begin{figure}
    \centering
    \input{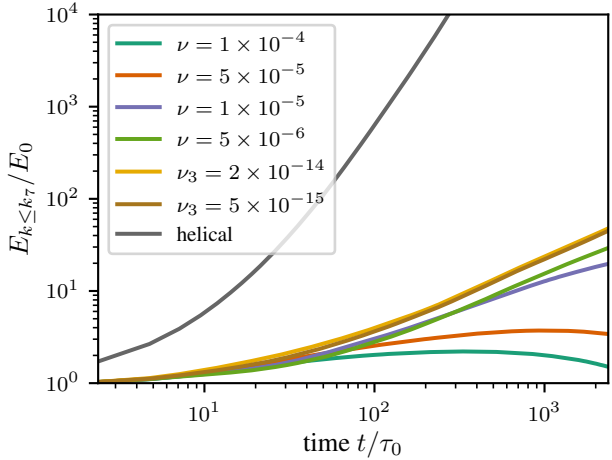}
    \caption{Time evolution of the energy on large scales for runs with different viscosities. Only certain parameters in the simulation setup give rise to the \textit{inverse transfer.}}
    \label{fig:visc_energy_increase}
\end{figure}
\begin{figure}
    \centering
    \input{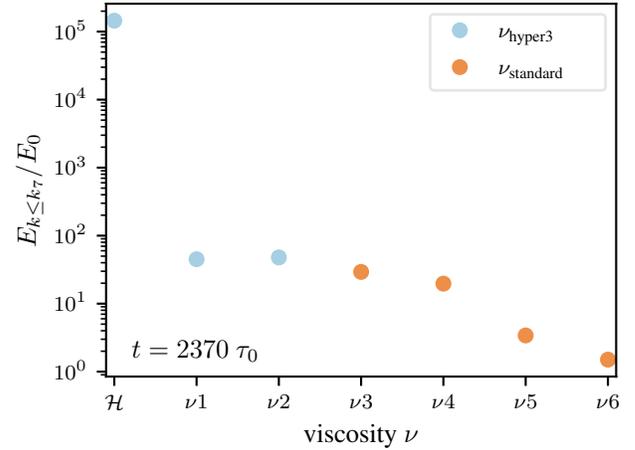}
   \caption{Comparison of the energy on large scales for the runs with different viscosities at time $t=100$. $\mathcal{H}$ denotes the helical comparison run. \hl{The values for the viscosities are given in} \cref{fig:viscosity_comparison}, where $\nu1$ is the upper left panel and $\nu6$ the panel in the lower right corner} 
    \label{fig:visc_t10_comp}
\end{figure}
We summarize the results of our viscosity study in \cref{fig:viscosity_comparison}, \hl{where we show the spectra at different times $t=0,20,2\e{2},1\e{3}\ \&\ 2\e{3}\ \tau_0$. }
We see that smaller viscosities, i.e. larger Reynolds numbers, lead to a stronger effect of the \textit{inverse transfer}.
If the Reynolds number drops below $500$ ($\nu\ge 5\e{-5}$), this effect is essentially not visible.
Otherwise, the effect is strongest when we use hyperviscosity. 

Note for small Laplacian viscosities the so-called bottleneck effect \citep[see][]{Dobler03} sets in, where energy is accumulated at the smallest scales, i.e. it does not dissipate. 
It looks like this small-scale effect does not impact the large scales.

In order to quantify the effect of the inverse transfer we measure the energy on large scales as a function of time as: 
\begin{equation}
E_{k\leq k_L}(t) =\int_0^{k_L} E_k(t)\ dk,
\end{equation}
where we choose $k_L = 7$ which is a scale not fully processed at the end of the simulation. 
We demonstrate the time evolution of the energy on large scales in \cref{fig:visc_energy_increase}, which also includes the maximally helical comparison run. 
Again, this analysis shows how the inverse transfer depends on the Reynolds number.
Nevertheless, this effect is much less efficient than an inverse cascade due to a helical magnetic field.

\Cref{fig:visc_t10_comp} \hl{shows $E_{k\leq k_L}/E_0$ at time $t= 2370\ \tau_0$.}
At this time the integrated large scale energy in the viscosity and hyperviscosity runs differ by a factor of a few, whereas the helical energy is already four orders of magnitude larger. 
\subsection{Prandtl Number Dependency}
\begin{figure*}
    \centering
    \input{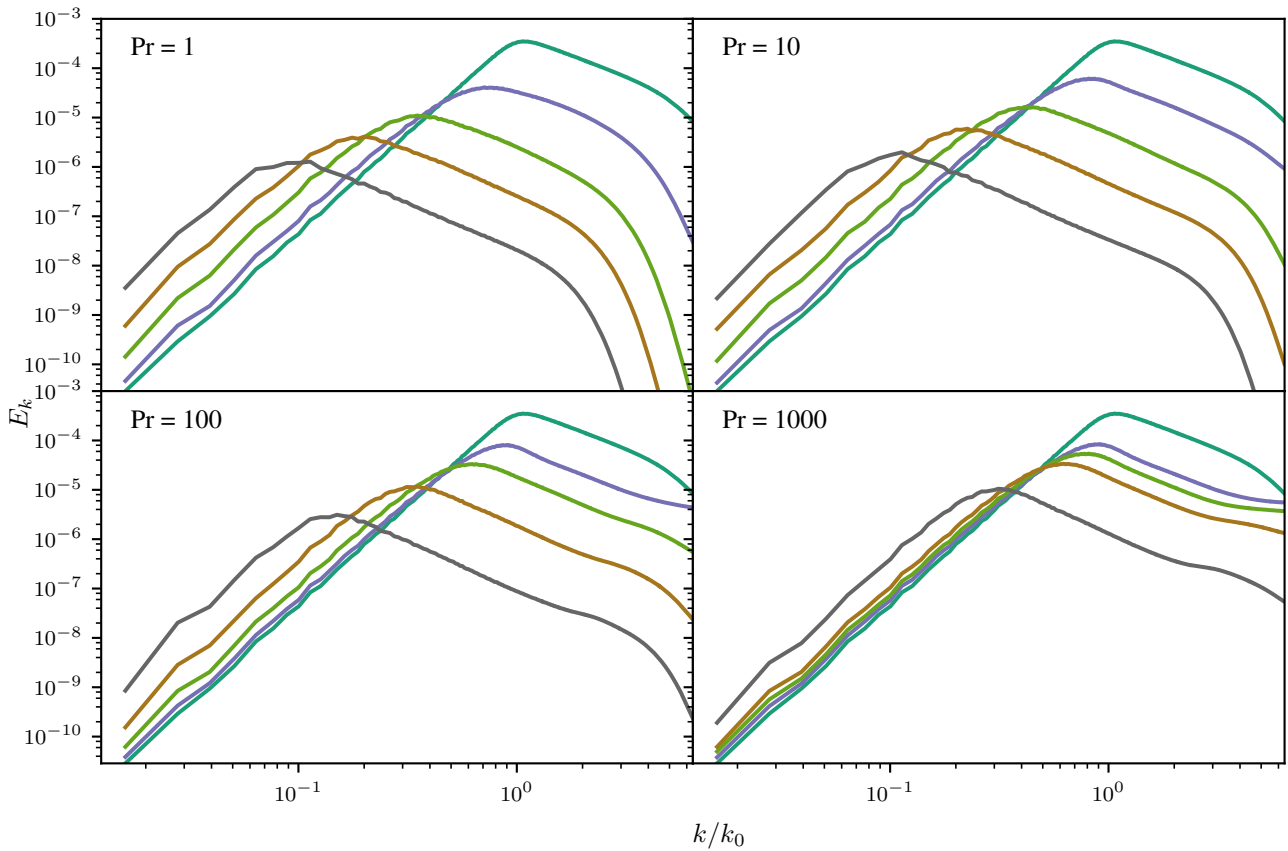}
    \caption{Magnetic power spectra of runs with different Prandtl numbers. \hl{The spectra are shown for $t=0,20,2\e{2},1\e{3},7\e{3} \tau_0$.}
    The effect of the \textit{inverse transfer} of energy is strongest with a Prandtl number of $\textrm{Pr}=1$ and becomes less pronounced for higher Prandtl numbers, 
    especially the run with $\textrm{Pr}=1000$ shows very little increase of energy on large scales.}
    \label{fig:prandtl_spectra_grid}
\end{figure*}
\begin{figure}
    \centering
    \input{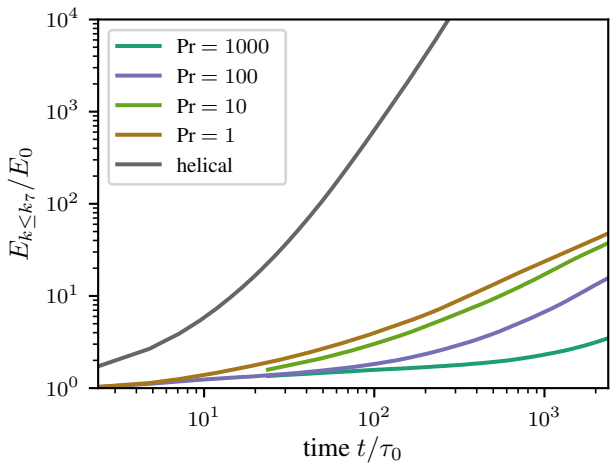}
    \caption{Time evolution of the energy on large scales. The transfer of energy to large scales depends crucially on the Prandtl number where larger Prandtl numbers show less efficient \textit{inverse transfer}. We also show the evolution of the energy on large scales for the helical case.}
    \label{fig:prandtl_energy_evolution}
\end{figure}
\begin{figure}
    \centering
    \input{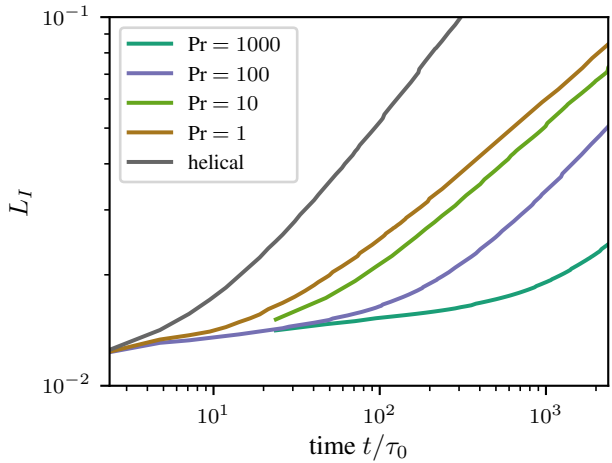}
    \caption{Evolution of the integral scale for runs with varying Prandtl numbers. Larger Prandtl numbers lead to a slower increase of the coherence length.}
    \label{fig:prandtl_scale_evolution}
\end{figure}
\begin{figure}
    \centering
    \input{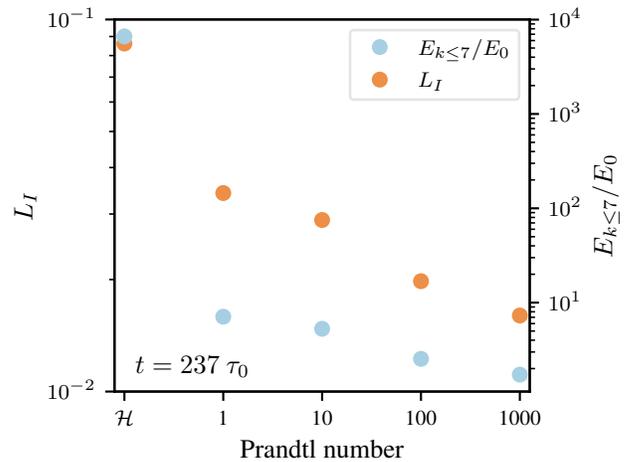}
    \caption{Comparison of the integral scale for the runs with different Prandtl numbers at the time $t=10$. $\mathcal{H}$ denotes the helical case.}
    \label{fig:prandtl_t10_comp}
\end{figure}
As mentioned earlier, we also studied the effect of the Prandtl number on the inverse transfer.
The Prandtl number is defined as the ratio between viscosity and magnetic diffusivity 
\begin{equation}
\textrm{Pr} = \nu_3 / \eta_3. 
\end{equation}
Where the index '3' indicates that we use hyperviscosity $\nu_3$ and hyperdiffusivity $\eta_3$. 
For the different runs we changed the diffusivity $\eta_3$ and kept the viscosity constant.
We vary $Pr$ from $1$ to $\textrm{Pr}=1000$.

As one can see in \cref{fig:prandtl_spectra_grid}, surprisingly, \textit{higher} Prandtl numbers slow down the inverse transfer. 
Especially in the run with the highest Prandtl number the effect of \textit{inverse transfer} ceases.
This is also quantified in \cref{fig:prandtl_energy_evolution} where we show the time evolution of the energy on large scales, with a comparison to the helical case. 
The strongest increase of magnetic energy on large scale is seen in the $Pr=1$ case. 
Furthermore, in \cref{fig:prandtl_scale_evolution} we show the evolution of the integral scale.
In \cref{fig:prandtl_t10_comp} we show the integral scale and the energy on large scales for our different runs \hl{at the time $t=200\ \tau_0$.}
This indicates again a clear trend of a weaker inverse transfer with increasing Prandtl number.
Again, none of the non-helical effects can compare with the \textit{inverse cascade} of the helical run.
\hl{A mechanism that could explain the transport of magnetic quantities is the merging of attracting magnetic flux densities with opposite sign, which was discussed by M\"uller (2012) \citep{Mueller12}. 
This does not act as a dynamo as it thins out the magnetic flux geometrically the larger the stuctures get. \\
Resistive MHD is essential for this reconnection process which can be visualised as two filaments with currents flowing in the same direction. With increasing magnetic Prandtl number at constant viscosity, as is the case in our simulations, this process becomes less and less effecient and explains our observations in our Prandtl number comparison runs. }\cite{MuellerPrivate}
\subsection{Back Reaction and Decay Laws}
   \begin{figure}
        \centering
        \input{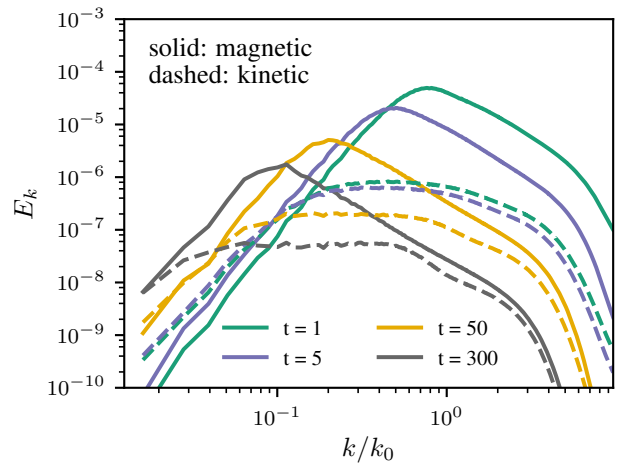}
         \caption{Magnetic and kinetic power spectra of the nonhelical simulation with a resolution of $N=1536^3$. Note at early times the kinetic part can exceed the magnetic part of the spectrum in large scales and might further excite fluctuations in the $B$-field.}
        \label{fig:nonhelical_mag_spec}
   \end{figure}
   \begin{figure}
        \centering
        \includegraphics[width=.95\linewidth]{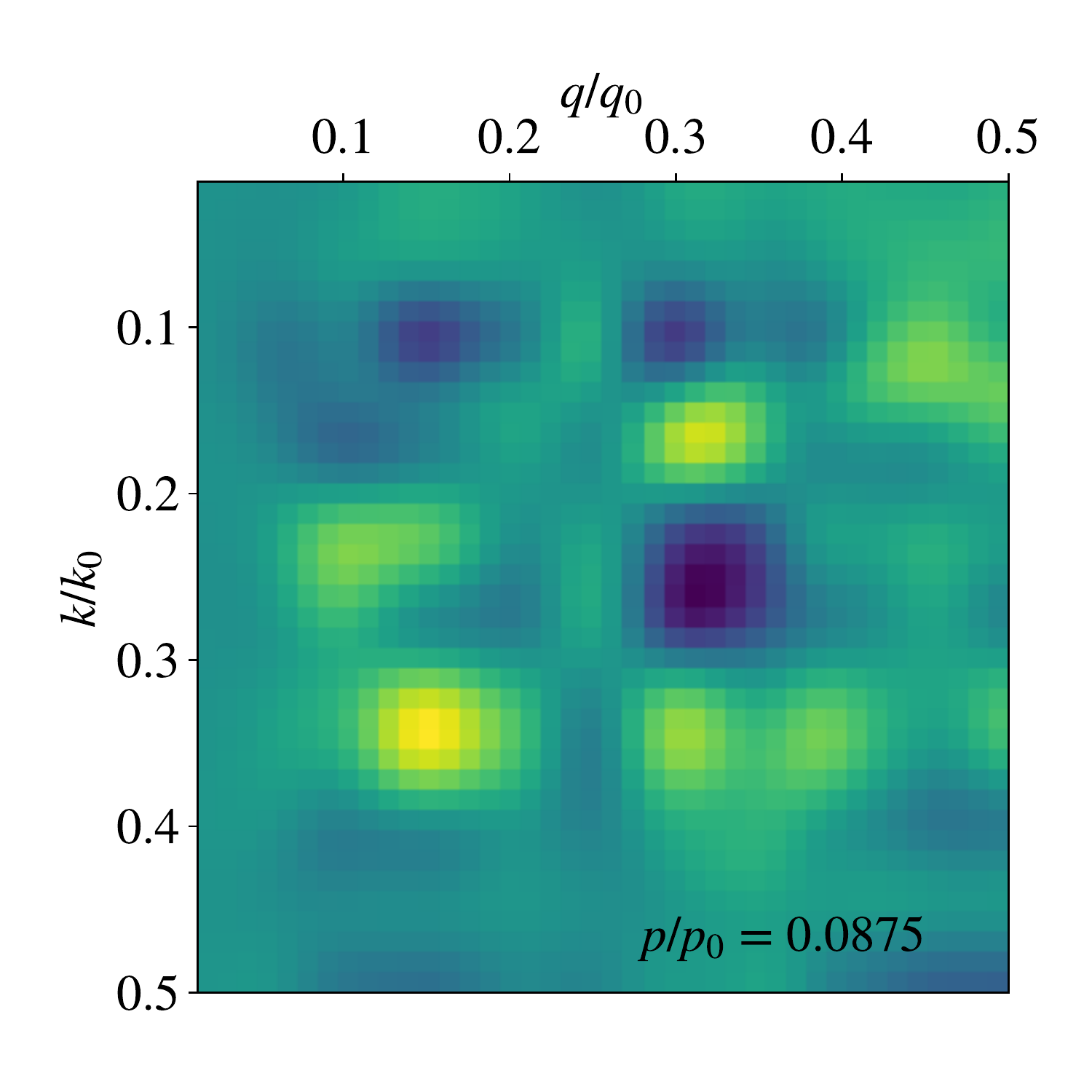}
         \caption{\hl{Spectral Transfer function $T_{kpq}$ as a function of $k$ \& $q$ at $p/k_0=0.0875$, following the analysis of Brandenburg (2001)\cite{Brandenburg01}. Yellow pixels indicate positive values and blue pixels negative values. See also} \cref{fig:nonhelical_mag_spec} \hl{for the time evolution of the magnetic energy spectrum}}
        \label{fig:transfer_function}
   \end{figure}
To study the the effect of possible back reactions from the velocity fluctuations on the magnetic field on scales beyond the integral scale, we performed a higher resolution run with $N=1536^3$ grid points.
Again, we initialize the velocity field with $\mathbf{u}=0$.

We chose the simulation parameters such that the effect of the inverse transfer of energy occurs most prominently, meaning low hyperviscosity parameter and a Prandtl number of $\textrm{Pr}=1$. 
We show the magnetic as well as the kinetic spectra of this run in \cref{fig:nonhelical_mag_spec}. 
As mentioned earlier, the fluctuations in the magnetic field excite velocity fluctuations to a strength until back reactions set in.
Generally, the power spectrum of the kinetic part has a different shape than the magnetic power spectrum. 
At early times the kinetic power spectrum exceeds the magnetic one on large scales, although this feature is not persistent.
Nevertheless, the energies continue on equipartition at scales way beyond the integral scale.
Since the spectrum of the velocity field is not confined to a steep $k^4$ spectrum, like the divergence-free magnetic field \citep[see][]{Durrer03}, it can go beyond the magnetic one on large scales.
In principle, this could be an explanation for the inverse transfer at early times.
On small scales, eventually, equipartition of both power spectra is reached.
On intermediate scale the kinetic power spectrum does not show a clear peak but rather a plateau, which never reaches the peak of the magnetic power spectrum.
This might be due to the intermittent structure of the magnetic field.
   \begin{figure}
        \centering
        \input{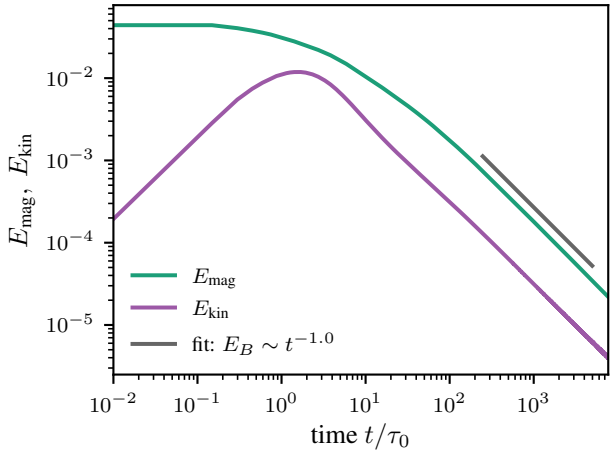}
         \caption{Evolution of total magnetic energy $E_\textrm{mag}$ and total kinetic energy $E_\textrm{kin}$. Starting from nonhelical initial condition with $k_{\textrm{peak}}= 80, N=1536^3$. The initial setup had zero velocity field, after a short relaxation time the kinetic energy decays as the magnetic energy} 
        \label{fig:nonhelical_ts}
   \end{figure}
   \begin{figure}
        \centering
        \input{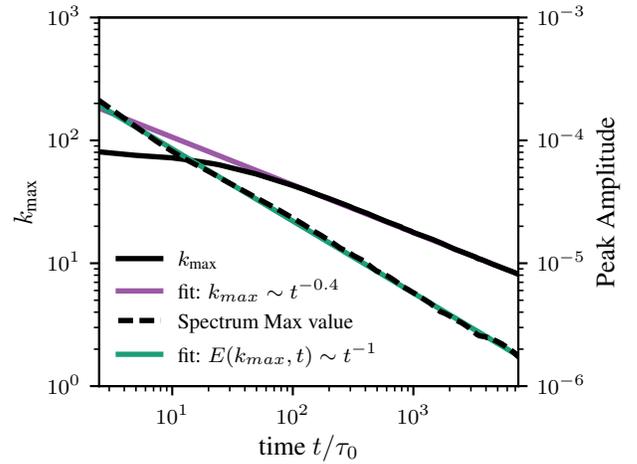}
         \caption{Time evolution of the peak of the spectrum shown in \cref{fig:nonhelical_mag_spec}. The solid black line shows the location in $k$-space of the spectrum's peak and the dashed line gives its energy value (right label). One can see that after an initial settling phase the peak shifts to lower $k$ values. Its peak energy follows the same law as the mean magnetic field energy shown in \cref{fig:nonhelical_ts}}
        \label{fig:nonhel_peak}
   \end{figure}

Nevertheless, as can be seen in \cref{fig:nonhelical_ts}, both energies evolve with a constant ratio of about $1/4$ and obey a decay law of $E\propto t^{-1}$. 
This decay law was also reported by Brandenburg et al. \citep{Brandenburg15} in the case of a strong \textit{inverse transfer} and is different from the expected value without inverse transfer which has a steeper decay law of $E\propto t^{-10/7}$.

\hl{A different way of examining the nature of the \textit{inverse transfer}, is to analyse the \textit{Spectral Transfer function} $T_{kpq} = \langle\mathbf{J}^k\cdot (\mathbf{u}^p\times \mathbf{B}^q)\rangle$.
In \cref{fig:transfer_function} we show one example,following Brandenburg (2001) \cite{Brandenburg01}. 
Indices $k,p,q$ indicate shells in Fourier space of our fields from the simulation at a late time in the run when the \textit{inverse transfer} is activ.
We show the 2-D plot for $k\ \&\ q$ for a fixed value of $p/k_0=7/80=0.0875$ in }\cref{fig:transfer_function}.
\hl{We see that the current $\mathbf{J}^k$ decreases and the magnetic field $\mathbf{B}^q$ increases on similar scales
as the MHD flux decreases. 
This transfer of energy via the MHD flux is mediated through the velocity field $\mathbf{u}_p$.
It occurs for small scales $p,k \lesssim 0.5\ k_0$.
This demonstrates the nature of the \textit{inverse transfer} as a large-scale process}

\Cref{fig:nonhel_peak} shows the evolution of the peak of the magnetic energy spectrum and its associated  wavenumber.  
Note that the peak amplitude starts to decay instantly, while the peak starts to shift only after a short initial phase. 
Again, this could be an indication that back reaction from the velocity field initiates the \textit{inverse transfer}.
\subsection{Delta Peak Energy Injection}
Furthermore, we run a simulation with a $\delta$-peak spectrum as initial conditions.
This setup is comparable to the initial conditions of the semi-analytic work of Saveliev et al \cite{Saveliev12}.  
In this study a $E\sim k^4$ spectrum develops self-consistently from those initial conditions. 
Note that in their calculations no more energy is transferred to large scales.

\begin{figure}
    \centering
    \input{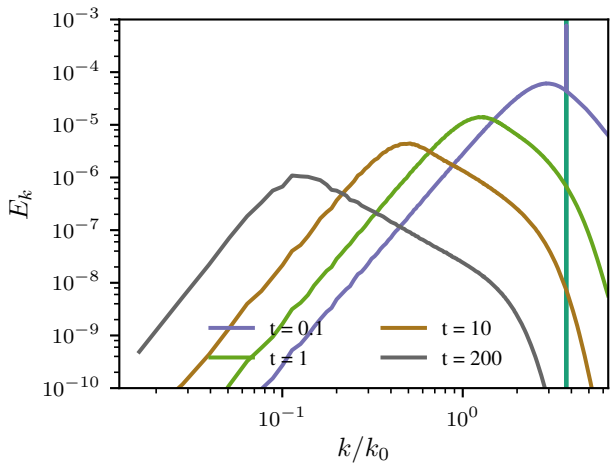}
    \caption{Magnetic power spectra of a simulation with energy injected at a single wavenumber $k=300$. An $E\sim k^4$ spectrum develops very quickly. After this redistribution of energy the spectrum evolves as in the nonhelical case with a continuous \textit{inverse transfer} of energy to large scales.} 
    \label{fig:injecting_k300}
\end{figure}

Here we choose the $\delta$-injection scale to be close to the Nyquist wavenumber $k_{Ny}=N/2$ so there is enough separation of simulation scales $k_0$ and $k_\delta$ in $k$-space. 
In \cref{fig:injecting_k300} we show the spectra of the simulation for $k_\delta = 300$. 
One can see that the expected \textit{causal} spectrum develops almost instantly. 
The spectrum has a peak which is at wavenumbers close to but less than the injecting wavenumber $k_\textrm{peak} < k_\delta$. 
For $k > k_\delta$ a turbulent spectrum with a Kolmogorov slope $E_k\propto k^{-5/3}$ develops. 
This resembles the initial conditions of the simulation previously performed in this analysis very closely. 
It is thus reasonable to assume our chosen initial conditions can also resemble the conditions after such a causal field generation process. 
After generating this turbulent spectrum the \textit{inverse transfer} of energy sets in and the energy carrying wavenumber $k_{I}$ decreases while the energy is decaying as in our previous simulations.

It has to be noted that in these initial conditions no helicity was injected into the fields explicitly. However, some, although negligible, helicity builds up over time. 
\subsection{Initial Slope Comparison}
\begin{figure*}
    \centering
    \input{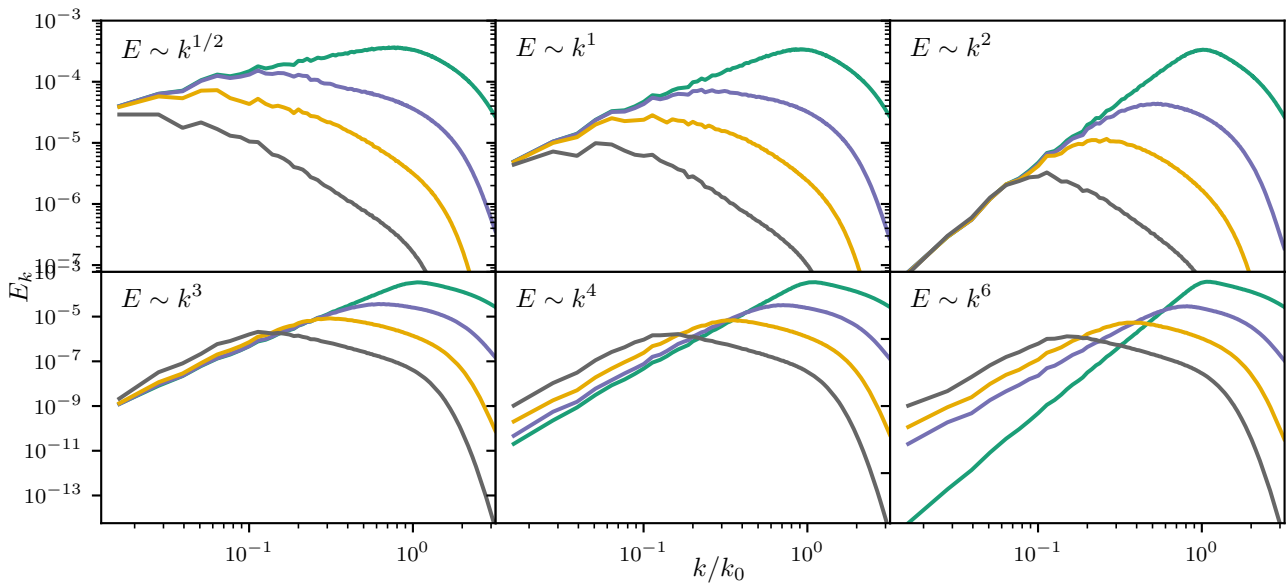}
    \caption{Magnetic power spectra of runs with different spectral indices $n$. The inverse transfer of energy can only be observed if there is a steep initial spectrum with $\ n\geq 3$. If an initial spectrum steeper than $k^4$ is given it will flatten to a \textit{causal} spectrum, similar to the $\delta$-peak run (see \cref{fig:injecting_k300}).}
    \label{fig:grid_spec_slope}
\end{figure*}
Finally, in our parameter study, we investigate the dependence of the \textit{inverse transfer} on the spectral index $n$ with a number of $512^3$ simulations.
 Here, we vary $n=0.5\dots 6$. 
We present the results of these initial conditions in \cref{fig:grid_spec_slope}. 
A steep magnetic field spectrum is needed in order to transfer energy from small to large scales during the decay.
Similar to the delta peak simulation, a $E\propto k^4$ spectrum builds up if the power spectrum is initially steeper than that.
This $n=4$ case is also the case where the inverse transfer is strongest.
A flatter spectrum shows a less efficient effect. 
For $n\leq 2$ the inverse transfer effect vanishes completely and the magnetic field only decays without significant pileup of energy on small scales. 
This is also a reason that earlier studies did not see the effect of the \textit{inverse transfer} \citep{Banerjee04b}.
We expect a more shallow spectrum of $E_k\sim k^{3/2}$, the so-called Kazantsev spectrum \cite{Kazantsev68}, fields generated by the small-scale dynamo \citep{Federrath11a, Schober12a, Wagstaff14}, so such a field would not go through an \textit{inverse transfer}, but only decay after its generation.
\subsection{Zeus-MP2 Comparison}
\begin{figure}
    \centering
    \input{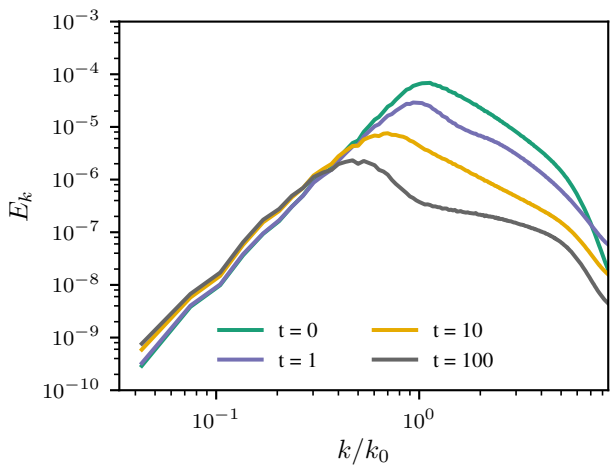}
    \caption{Zeus-MP2 run with Pencil-Code initial conditions. With the Zeus Code there is a little increase of power at low k, indicating it does not have high enough Reynolds numbers.} 
    \label{fig:zeus_pencil_init}
\end{figure}

We also ran one simulation with the \textit{Zeus-MP2} code in order to compare the results to the \textsc{Pencil-Code}. 
The Zeus code employs only a second order finite difference scheme to integrate the MHD equation and hence is more dissipative than the \textsc{Pencil-Code}.
This is reflected in \cref{fig:zeus_pencil_init} where we show the magnetic power spectrum.
In particular the inertial range is not as pronounced as in the runs with hyperviscosity with the pencil-code.
Additionally, due to the effectively smaller Reynolds number in the Zeus run, the effect of the \textit{inverse transfer} is not observed. 
Note also that the integral scale at no times moves to scales larger than the scales imprinted by the initial blue spectrum.
Another difference in the setup is the location of the peak in the spectrum, shown in \cref{fig:zeus_pencil_init}. We have discovered that to observe an \textit{inverse cascade} like effect, the main energy carrying scale must be at high $k$. 
This caused numerical problems with the code if too much power resided at large wavenumbers $k$, limiting the scale for the peak of the spectrum.
\section{Conclusions}
\label{sec:conclusions}
In this work we presented a parameter study based on high-resolution numerical simulations of decaying MHD turbulence. 
We explored a wide range of numerical parameters and initial conditions in order to find a pattern at which the \textit{inverse transfer} of magnetic energy from small scales to large scales takes place for nonhelical magnetic fields. 

Our most prominent finding is the surprising dependency on the Prandtl number: Larger Prandtl numbers lead to a less efficient \textit{inverse transfer} of magnetic energy and might be fully suppressed at Prandtl numbers larger than $10^3$.
This raises the question whether one can apply the effect of the \textit{inverse transfer} of energy to the evolution of magnetic fields in the early Universe. 
There, one expects large Prandtl numbers of $\mathrm{Pr} \sim 10^8(T/\mathrm{keV})^{-3/2}$ \citep{Banerjee04b}.
For instance, considering a causally generated field, it will decay according to $E\sim t^{-10/7}$ in the case of suppressed inverse transfer.
It will decay as $E\sim t^{-1}$, however, if the inverse transfer is efficient.
This results in a many orders of magnitudes weaker field in the former case.
Therefore, it is questionable whether EW phase transition generated fields could be significant enough today to account for the assumed fields in the voids of galaxies  (see \cite{Wagstaff16} and \citep{Kahniashvili13}).

Furthermore, the efficiency of the \textit{inverse transfer} depends on the Reynolds number. 
Here, the Reynolds number has to be sufficiently large to observe the effect.
With our Pencil-Code simulations we find a critical Reynolds number of $\mathrm{Re}=500$ for a Prandtl number of $\mathrm{Pr}=1$.

Another very interesting result of our study is that for shallow and moderately steep slopes of the magnetic power spectrum $n\leq 2$ the effect of the \textit{inverse transfer} is not present.
Again, this dependence on $n$ is qualitatively different than the \textit{inverse cascade} of helical fields, which is independent of the spectral index $n$. 
An $n=2$ case could be expected from an average over a stochastic distribution of magnetic dipoles \cite{Hogan83}, and a field with $n=3/2$ will be generated by the small-scale dynamo \cite{Kazantsev68}. 

Although we find numerical evidence for the effect of non-helical \textit{inverse transfer}, the physics behind this mechanism is yet to be determined.
One option could be the enhancement of the magnetic field on large scales by back reactions of velocity fluctuations where the kinetic power spectrum can exceed the magnetic one on scales above the correlation length.
Nevertheless, our simulations do not indicate that this mechanism could persist throughout the entire decay phase. See also \cite{Brandenburg15}.

Additionally, Brandenburg et al \cite{Brandenburg15} suspect an effect of two-dimensional structure of the turbulence. 
In two dimensions the square of the vector potential $\langle \mathbf{A}^2 \rangle$ is conserved \citep{BiskampBook} and could serve as an explanation to the non-helical \textit{inverse transfer}, similar to the conserved helicity in three dimensions.
We show the evolution of $\langle \mathbf{A}^2 \rangle$  in \cref{fig:arms_ts} where one sees that it varies by at least $80\%$. This is not a lot compared to the magnetic energy which changes by three orders of magnitude during its decay.
On the other hand, the helicity in the maximally helical run changes only by a few percent.
Nevertheless, the conservation of $\langle \mathbf{A}^2\rangle$  is not expected in three dimensions and it is not clear why a two-dimensional turbulent structure should develop (depending on the Prandtl number and initial spectral index).

\hl{M\"uller} \hl{suggests}, \cite{Mueller12, MuellerPrivate} \hl{ the \textit{inverse transfer} could be an effect of merging current densities. Especially in the resistive MHD case this seems to be a viable possibility, since it can explain the difference in behaviour at large Prandtl numbers, where the merging process becomes inefficient.
We do not have any quantification of how important this merger is to explain the \textit{inverse transfer}, though.}

Another way of explaining the inverse transfer is the assumption of a self-similar evolution of the decaying MHD turbulence.
Using rescaled MHD variables, Olesen \cite{Olesen97} and \citep{Campanelli2015} constructed such a self-similar scenario of decaying MHD turbulence.
\hl{Although the rescaling of the MHD variables is generally not restricted to a specific choice of the rescaling function, an \textit{inverse transfer} can only be explained by a very specific one where the viscosity is not rescaled, i.e. $\nu\rightarrow l^0\nu$ and $l$ is the scale function.
First of all, there is no physical reasoning for this (unmotivated) choice and furthermore the rescaling of variables should not impact the physical result.
The specific self-similar solution resulting in the \textit{inverse transfer} does not give further insight to this problem. }

We thus conclude that while there is numerical evidence from our simulations that the non-helical \textit{inverse transfer} of energy can be present, a satisfying physical explanation is still missing. 
\begin{acknowledgments}
      Part of this work was supported by the German
      \emph{Deut\-sche For\-schungs\-ge\-mein\-schaft} (DFG), Sonderforschungsbereich 676 section C9.
      We ran the code on the \textsc{Hummel Cluster} at the \textit{Regionales Rechenzentrum} of the university of Hamburg. Special thanks also go to Jacques Wagstaff and Pranjal Trivedi for their help and useful discussions. We thank NORDITA for the hospitality during the workshop \textit{Origin, Evolution, and Signatures of Cosmological Magnetic fields} in 2015.
\end{acknowledgments}
\appendix
\section{Helicity in the simulation}
\begin{figure}
  \centering
  \input{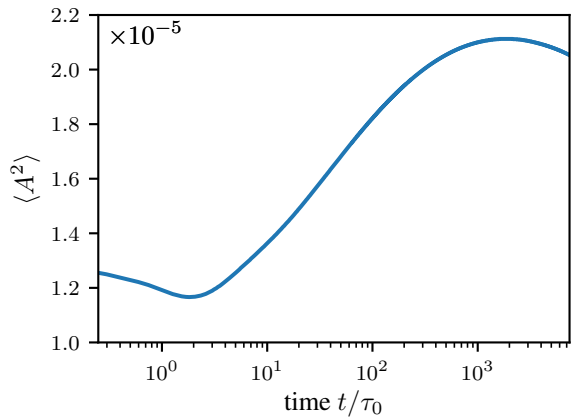}
  \caption{Temporal evolution of the square of the magnetic vector potential.}
  \label{fig:arms_ts}
\end{figure}
\begin{figure}
 \centering
 \input{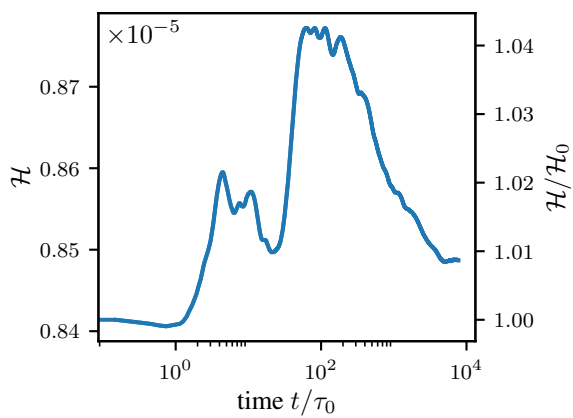}
 \caption{Temporal evolution of the total helicity for the large non-helical run. The magnitude of the helicity is of the order $1\e{-5}$ compared to $\mathcal{H}_\textrm{max}=0.32$ in the maximally helical case.}
 \label{fig:helicity_ts}
\end{figure}
\begin{figure}
 \centering
 \input{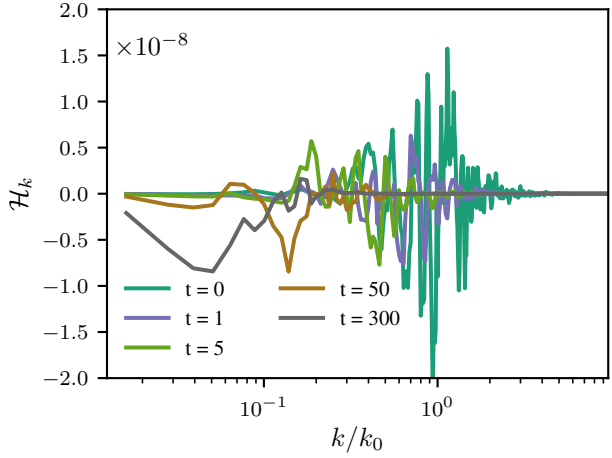}
 \caption{Spectrum of helicity fluctuations $\mathcal{H}_k$.}
 \label{fig:helicity_spec}
\end{figure}
\begin{figure}
 \centering
 \input{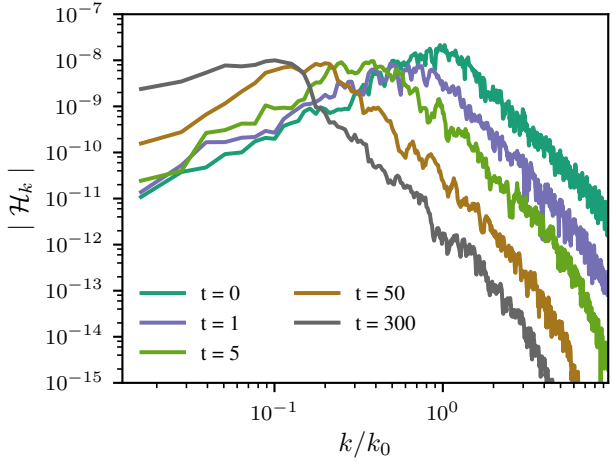}
 \caption{Spectrum of absolute value of helicity fluctuations $\mathcal{H}_k$. A similar behaviour as in the energy spectrum can be seen. For the relative fluctuations see \cref{fig:helicity_spec_ratio}}
 \label{fig:helicity_spec_abs}
\end{figure}
\begin{figure}
 \centering
 \input{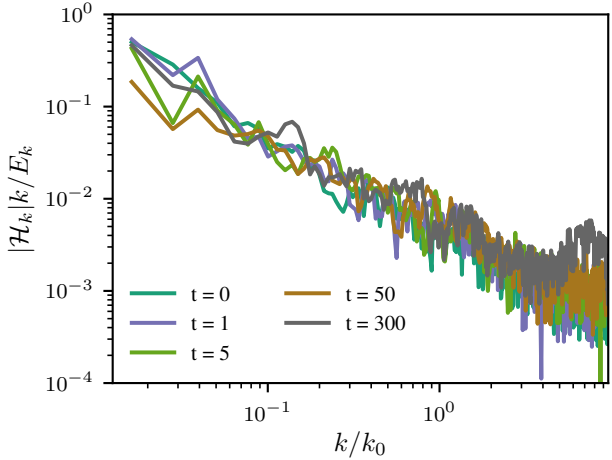}
 \caption{Relative error of helicity fluctuations using the data from \cref{fig:nonhelical_mag_spec,fig:helicity_spec_abs}.}
 \label{fig:helicity_spec_ratio}
\end{figure}

In order to check whether an artificial build up of helicity might influence the magnetic field evolution, we analyze the numerical helicity in the simulation of our non-helical high resolution run.
In \cref{fig:helicity_ts} we show the time evolution of the total helicity which is of the order of $\mathcal{H}\sim 10^{-5}$.
This corresponds roughly to the numerical helicity error (note in the maximal helical case the total helicity is about $\mathcal{H}={0.32}$).
Furthermore, we show a spectral analysis of the helicity in \cref{fig:helicity_spec,fig:helicity_spec_abs}.
The former one shows the actual helicity spectra, whereas the latter shows its absolute value which, surprisingly, shows features similar to the magnetic power spectra.
It is surprising since those fluctuations should be purely numerical and should not trace physical properties.
Additionally, we compute the error from the helicity fluctuations $E_k/\mathcal{H}_kk$ in our simulation which we show in \cref{fig:helicity_spec_ratio}.
\bibliography{astro_johannes}
\end{document}

%% file: journals.tex
\def\jnl@style#1{{\rmfamily#1}}%
\def\jref@jnl#1{{\jnl@style#1}}%

\newcommand\aj{\jref@jnl{AJ}}%
\newcommand\rmxaa{\jref@jnl{RMXAA}}%
\newcommand\araa{\jref@jnl{ARA\&A}}%
\newcommand\apjl{\jref@jnl{ApJ}}%
\newcommand\apjs{\jref@jnl{ApJS}}%
\newcommand\apss{\jref@jnl{Ap\&SS}}%
\newcommand\aap{\jref@jnl{A\&A}}%
\newcommand\aapr{\jref@jnl{A\&A~Rev.}}%
\newcommand\aaps{\jref@jnl{A\&AS}}%
\newcommand\azh{\jref@jnl{AZh}}%
\newcommand\baas{\jref@jnl{BAAS}}%
\newcommand\jrasc{\jref@jnl{JRASC}}%
\newcommand\memras{\jref@jnl{MmRAS}}%
\newcommand\mnras{\jref@jnl{MNRAS}}%
\newcommand\na{\jref@jnl{New Astron.}}%
\newcommand\nar{\jref@jnl{New Astron. Rev.}}%
\newcommand\pasp{\jref@jnl{PASP}}%
\newcommand\pasj{\jref@jnl{PASJ}}%
\newcommand\qjras{\jref@jnl{QJRAS}}%
\newcommand\skytel{\jref@jnl{S\&T}}%
\newcommand\solphys{\jref@jnl{Sol.~Phys.}}%
\newcommand\sovast{\jref@jnl{Soviet~Ast.}}%
\newcommand\ssr{\jref@jnl{Space~Sci.~Rev.}}%
\newcommand\zap{\jref@jnl{ZAp}}%
\newcommand\iaucirc{\jref@jnl{IAU~Circ.}}%
\newcommand\aplett{\jref@jnl{Astrophys.~Lett.}}%
\newcommand\apspr{\jref@jnl{Astrophys.~Space~Phys.~Res.}}%
\newcommand\bain{\jref@jnl{Bull.~Astron.~Inst.~Netherlands}}%
\newcommand\fcp{\jref@jnl{Fund.~Cosmic~Phys.}}%
\newcommand\gca{\jref@jnl{Geochim.~Cosmochim.~Acta}}%
\newcommand\grl{\jref@jnl{Geophys.~Res.~Lett.}}%
\newcommand\jcap{\jref@jnl{J.~Cosmology.~Astroparticle.~Phys.}}
\newcommand\jgr{\jref@jnl{J.~Geophys.~Res.}}%
\newcommand\jqsrt{\jref@jnl{J.~Quant.~Spec.~Radiat.~Transf.}}%
\newcommand\memsai{\jref@jnl{Mem.~Soc.~Astron.~Italiana}}%
\newcommand\nphysa{\jref@jnl{Nucl.~Phys.~A}}%
\newcommand\physrep{\jref@jnl{Phys.~Rep.}}%
\newcommand\physscr{\jref@jnl{Phys.~Scr}}%
\newcommand\planss{\jref@jnl{Planet.~Space~Sci.}}%
\newcommand\procspie{\jref@jnl{Proc.~SPIE}}%

%% file: fitting_example_nonhel_k80.pgf
\begingroup%
\makeatletter%
\begin{pgfpicture}%
\pgfpathrectangle{\pgfpointorigin}{\pgfqpoint{3.259842in}{2.444882in}}%
\pgfusepath{use as bounding box, clip}%
\begin{pgfscope}%
\pgfsetbuttcap%
\pgfsetmiterjoin%
\definecolor{currentfill}{rgb}{1.000000,1.000000,1.000000}%
\pgfsetfillcolor{currentfill}%
\pgfsetlinewidth{0.000000pt}%
\definecolor{currentstroke}{rgb}{1.000000,1.000000,1.000000}%
\pgfsetstrokecolor{currentstroke}%
\pgfsetdash{}{0pt}%
\pgfpathmoveto{\pgfqpoint{0.000000in}{0.000000in}}%
\pgfpathlineto{\pgfqpoint{3.259842in}{0.000000in}}%
\pgfpathlineto{\pgfqpoint{3.259842in}{2.444882in}}%
\pgfpathlineto{\pgfqpoint{0.000000in}{2.444882in}}%
\pgfpathclose%
\pgfusepath{fill}%
\end{pgfscope}%
\begin{pgfscope}%
\pgfsetbuttcap%
\pgfsetmiterjoin%
\definecolor{currentfill}{rgb}{1.000000,1.000000,1.000000}%
\pgfsetfillcolor{currentfill}%
\pgfsetlinewidth{0.000000pt}%
\definecolor{currentstroke}{rgb}{0.000000,0.000000,0.000000}%
\pgfsetstrokecolor{currentstroke}%
\pgfsetstrokeopacity{0.000000}%
\pgfsetdash{}{0pt}%
\pgfpathmoveto{\pgfqpoint{0.573273in}{0.415224in}}%
\pgfpathlineto{\pgfqpoint{3.159147in}{0.415224in}}%
\pgfpathlineto{\pgfqpoint{3.159147in}{2.368215in}}%
\pgfpathlineto{\pgfqpoint{0.573273in}{2.368215in}}%
\pgfpathclose%
\pgfusepath{fill}%
\end{pgfscope}%
\begin{pgfscope}%
\pgfsetbuttcap%
\pgfsetroundjoin%
\definecolor{currentfill}{rgb}{0.000000,0.000000,0.000000}%
\pgfsetfillcolor{currentfill}%
\pgfsetlinewidth{0.803000pt}%
\definecolor{currentstroke}{rgb}{0.000000,0.000000,0.000000}%
\pgfsetstrokecolor{currentstroke}%
\pgfsetdash{}{0pt}%
\pgfsys@defobject{currentmarker}{\pgfqpoint{0.000000in}{-0.048611in}}{\pgfqpoint{0.000000in}{0.000000in}}{%
\pgfpathmoveto{\pgfqpoint{0.000000in}{0.000000in}}%
\pgfpathlineto{\pgfqpoint{0.000000in}{-0.048611in}}%
\pgfusepath{stroke,fill}%
}%
\begin{pgfscope}%
\pgfsys@transformshift{0.573273in}{0.415224in}%
\pgfsys@useobject{currentmarker}{}%
\end{pgfscope}%
\end{pgfscope}%
\begin{pgfscope}%
\pgftext[x=0.573273in,y=0.318002in,,top]{\rmfamily\fontsize{8.000000}{9.600000}\selectfont \(\displaystyle 10\)}%
\end{pgfscope}%
\begin{pgfscope}%
\pgfsetbuttcap%
\pgfsetroundjoin%
\definecolor{currentfill}{rgb}{0.000000,0.000000,0.000000}%
\pgfsetfillcolor{currentfill}%
\pgfsetlinewidth{0.602250pt}%
\definecolor{currentstroke}{rgb}{0.000000,0.000000,0.000000}%
\pgfsetstrokecolor{currentstroke}%
\pgfsetdash{}{0pt}%
\pgfsys@defobject{currentmarker}{\pgfqpoint{0.000000in}{-0.027778in}}{\pgfqpoint{0.000000in}{0.000000in}}{%
\pgfpathmoveto{\pgfqpoint{0.000000in}{0.000000in}}%
\pgfpathlineto{\pgfqpoint{0.000000in}{-0.027778in}}%
\pgfusepath{stroke,fill}%
}%
\begin{pgfscope}%
\pgfsys@transformshift{1.686948in}{0.415224in}%
\pgfsys@useobject{currentmarker}{}%
\end{pgfscope}%
\end{pgfscope}%
\begin{pgfscope}%
\pgftext[x=1.686948in,y=0.340224in,,top]{\rmfamily\fontsize{8.000000}{9.600000}\selectfont \(\displaystyle 20\)}%
\end{pgfscope}%
\begin{pgfscope}%
\pgfsetbuttcap%
\pgfsetroundjoin%
\definecolor{currentfill}{rgb}{0.000000,0.000000,0.000000}%
\pgfsetfillcolor{currentfill}%
\pgfsetlinewidth{0.602250pt}%
\definecolor{currentstroke}{rgb}{0.000000,0.000000,0.000000}%
\pgfsetstrokecolor{currentstroke}%
\pgfsetdash{}{0pt}%
\pgfsys@defobject{currentmarker}{\pgfqpoint{0.000000in}{-0.027778in}}{\pgfqpoint{0.000000in}{0.000000in}}{%
\pgfpathmoveto{\pgfqpoint{0.000000in}{0.000000in}}%
\pgfpathlineto{\pgfqpoint{0.000000in}{-0.027778in}}%
\pgfusepath{stroke,fill}%
}%
\begin{pgfscope}%
\pgfsys@transformshift{2.338407in}{0.415224in}%
\pgfsys@useobject{currentmarker}{}%
\end{pgfscope}%
\end{pgfscope}%
\begin{pgfscope}%
\pgftext[x=2.338407in,y=0.340224in,,top]{\rmfamily\fontsize{8.000000}{9.600000}\selectfont \(\displaystyle 30\)}%
\end{pgfscope}%
\begin{pgfscope}%
\pgfsetbuttcap%
\pgfsetroundjoin%
\definecolor{currentfill}{rgb}{0.000000,0.000000,0.000000}%
\pgfsetfillcolor{currentfill}%
\pgfsetlinewidth{0.602250pt}%
\definecolor{currentstroke}{rgb}{0.000000,0.000000,0.000000}%
\pgfsetstrokecolor{currentstroke}%
\pgfsetdash{}{0pt}%
\pgfsys@defobject{currentmarker}{\pgfqpoint{0.000000in}{-0.027778in}}{\pgfqpoint{0.000000in}{0.000000in}}{%
\pgfpathmoveto{\pgfqpoint{0.000000in}{0.000000in}}%
\pgfpathlineto{\pgfqpoint{0.000000in}{-0.027778in}}%
\pgfusepath{stroke,fill}%
}%
\begin{pgfscope}%
\pgfsys@transformshift{2.800624in}{0.415224in}%
\pgfsys@useobject{currentmarker}{}%
\end{pgfscope}%
\end{pgfscope}%
\begin{pgfscope}%
\pgftext[x=2.800624in,y=0.340224in,,top]{\rmfamily\fontsize{8.000000}{9.600000}\selectfont \(\displaystyle 40\)}%
\end{pgfscope}%
\begin{pgfscope}%
\pgfsetbuttcap%
\pgfsetroundjoin%
\definecolor{currentfill}{rgb}{0.000000,0.000000,0.000000}%
\pgfsetfillcolor{currentfill}%
\pgfsetlinewidth{0.602250pt}%
\definecolor{currentstroke}{rgb}{0.000000,0.000000,0.000000}%
\pgfsetstrokecolor{currentstroke}%
\pgfsetdash{}{0pt}%
\pgfsys@defobject{currentmarker}{\pgfqpoint{0.000000in}{-0.027778in}}{\pgfqpoint{0.000000in}{0.000000in}}{%
\pgfpathmoveto{\pgfqpoint{0.000000in}{0.000000in}}%
\pgfpathlineto{\pgfqpoint{0.000000in}{-0.027778in}}%
\pgfusepath{stroke,fill}%
}%
\begin{pgfscope}%
\pgfsys@transformshift{3.159147in}{0.415224in}%
\pgfsys@useobject{currentmarker}{}%
\end{pgfscope}%
\end{pgfscope}%
\begin{pgfscope}%
\pgftext[x=3.159147in,y=0.340224in,,top]{\rmfamily\fontsize{8.000000}{9.600000}\selectfont \(\displaystyle 50\)}%
\end{pgfscope}%
\begin{pgfscope}%
\pgftext[x=1.866210in,y=0.164322in,,top]{\rmfamily\fontsize{10.000000}{12.000000}\selectfont \(\displaystyle k\)}%
\end{pgfscope}%
\begin{pgfscope}%
\pgfsetbuttcap%
\pgfsetroundjoin%
\definecolor{currentfill}{rgb}{0.000000,0.000000,0.000000}%
\pgfsetfillcolor{currentfill}%
\pgfsetlinewidth{0.803000pt}%
\definecolor{currentstroke}{rgb}{0.000000,0.000000,0.000000}%
\pgfsetstrokecolor{currentstroke}%
\pgfsetdash{}{0pt}%
\pgfsys@defobject{currentmarker}{\pgfqpoint{-0.048611in}{0.000000in}}{\pgfqpoint{0.000000in}{0.000000in}}{%
\pgfpathmoveto{\pgfqpoint{0.000000in}{0.000000in}}%
\pgfpathlineto{\pgfqpoint{-0.048611in}{0.000000in}}%
\pgfusepath{stroke,fill}%
}%
\begin{pgfscope}%
\pgfsys@transformshift{0.573273in}{0.415224in}%
\pgfsys@useobject{currentmarker}{}%
\end{pgfscope}%
\end{pgfscope}%
\begin{pgfscope}%
\pgftext[x=0.219877in,y=0.376093in,left,base]{\rmfamily\fontsize{8.000000}{9.600000}\selectfont \(\displaystyle 10^{-6}\)}%
\end{pgfscope}%
\begin{pgfscope}%
\pgfsetbuttcap%
\pgfsetroundjoin%
\definecolor{currentfill}{rgb}{0.000000,0.000000,0.000000}%
\pgfsetfillcolor{currentfill}%
\pgfsetlinewidth{0.803000pt}%
\definecolor{currentstroke}{rgb}{0.000000,0.000000,0.000000}%
\pgfsetstrokecolor{currentstroke}%
\pgfsetdash{}{0pt}%
\pgfsys@defobject{currentmarker}{\pgfqpoint{-0.048611in}{0.000000in}}{\pgfqpoint{0.000000in}{0.000000in}}{%
\pgfpathmoveto{\pgfqpoint{0.000000in}{0.000000in}}%
\pgfpathlineto{\pgfqpoint{-0.048611in}{0.000000in}}%
\pgfusepath{stroke,fill}%
}%
\begin{pgfscope}%
\pgfsys@transformshift{0.573273in}{2.142288in}%
\pgfsys@useobject{currentmarker}{}%
\end{pgfscope}%
\end{pgfscope}%
\begin{pgfscope}%
\pgftext[x=0.219877in,y=2.103157in,left,base]{\rmfamily\fontsize{8.000000}{9.600000}\selectfont \(\displaystyle 10^{-5}\)}%
\end{pgfscope}%
\begin{pgfscope}%
\pgfsetbuttcap%
\pgfsetroundjoin%
\definecolor{currentfill}{rgb}{0.000000,0.000000,0.000000}%
\pgfsetfillcolor{currentfill}%
\pgfsetlinewidth{0.602250pt}%
\definecolor{currentstroke}{rgb}{0.000000,0.000000,0.000000}%
\pgfsetstrokecolor{currentstroke}%
\pgfsetdash{}{0pt}%
\pgfsys@defobject{currentmarker}{\pgfqpoint{-0.027778in}{0.000000in}}{\pgfqpoint{0.000000in}{0.000000in}}{%
\pgfpathmoveto{\pgfqpoint{0.000000in}{0.000000in}}%
\pgfpathlineto{\pgfqpoint{-0.027778in}{0.000000in}}%
\pgfusepath{stroke,fill}%
}%
\begin{pgfscope}%
\pgfsys@transformshift{0.573273in}{0.935122in}%
\pgfsys@useobject{currentmarker}{}%
\end{pgfscope}%
\end{pgfscope}%
\begin{pgfscope}%
\pgfsetbuttcap%
\pgfsetroundjoin%
\definecolor{currentfill}{rgb}{0.000000,0.000000,0.000000}%
\pgfsetfillcolor{currentfill}%
\pgfsetlinewidth{0.602250pt}%
\definecolor{currentstroke}{rgb}{0.000000,0.000000,0.000000}%
\pgfsetstrokecolor{currentstroke}%
\pgfsetdash{}{0pt}%
\pgfsys@defobject{currentmarker}{\pgfqpoint{-0.027778in}{0.000000in}}{\pgfqpoint{0.000000in}{0.000000in}}{%
\pgfpathmoveto{\pgfqpoint{0.000000in}{0.000000in}}%
\pgfpathlineto{\pgfqpoint{-0.027778in}{0.000000in}}%
\pgfusepath{stroke,fill}%
}%
\begin{pgfscope}%
\pgfsys@transformshift{0.573273in}{1.239243in}%
\pgfsys@useobject{currentmarker}{}%
\end{pgfscope}%
\end{pgfscope}%
\begin{pgfscope}%
\pgfsetbuttcap%
\pgfsetroundjoin%
\definecolor{currentfill}{rgb}{0.000000,0.000000,0.000000}%
\pgfsetfillcolor{currentfill}%
\pgfsetlinewidth{0.602250pt}%
\definecolor{currentstroke}{rgb}{0.000000,0.000000,0.000000}%
\pgfsetstrokecolor{currentstroke}%
\pgfsetdash{}{0pt}%
\pgfsys@defobject{currentmarker}{\pgfqpoint{-0.027778in}{0.000000in}}{\pgfqpoint{0.000000in}{0.000000in}}{%
\pgfpathmoveto{\pgfqpoint{0.000000in}{0.000000in}}%
\pgfpathlineto{\pgfqpoint{-0.027778in}{0.000000in}}%
\pgfusepath{stroke,fill}%
}%
\begin{pgfscope}%
\pgfsys@transformshift{0.573273in}{1.455020in}%
\pgfsys@useobject{currentmarker}{}%
\end{pgfscope}%
\end{pgfscope}%
\begin{pgfscope}%
\pgfsetbuttcap%
\pgfsetroundjoin%
\definecolor{currentfill}{rgb}{0.000000,0.000000,0.000000}%
\pgfsetfillcolor{currentfill}%
\pgfsetlinewidth{0.602250pt}%
\definecolor{currentstroke}{rgb}{0.000000,0.000000,0.000000}%
\pgfsetstrokecolor{currentstroke}%
\pgfsetdash{}{0pt}%
\pgfsys@defobject{currentmarker}{\pgfqpoint{-0.027778in}{0.000000in}}{\pgfqpoint{0.000000in}{0.000000in}}{%
\pgfpathmoveto{\pgfqpoint{0.000000in}{0.000000in}}%
\pgfpathlineto{\pgfqpoint{-0.027778in}{0.000000in}}%
\pgfusepath{stroke,fill}%
}%
\begin{pgfscope}%
\pgfsys@transformshift{0.573273in}{1.622390in}%
\pgfsys@useobject{currentmarker}{}%
\end{pgfscope}%
\end{pgfscope}%
\begin{pgfscope}%
\pgfsetbuttcap%
\pgfsetroundjoin%
\definecolor{currentfill}{rgb}{0.000000,0.000000,0.000000}%
\pgfsetfillcolor{currentfill}%
\pgfsetlinewidth{0.602250pt}%
\definecolor{currentstroke}{rgb}{0.000000,0.000000,0.000000}%
\pgfsetstrokecolor{currentstroke}%
\pgfsetdash{}{0pt}%
\pgfsys@defobject{currentmarker}{\pgfqpoint{-0.027778in}{0.000000in}}{\pgfqpoint{0.000000in}{0.000000in}}{%
\pgfpathmoveto{\pgfqpoint{0.000000in}{0.000000in}}%
\pgfpathlineto{\pgfqpoint{-0.027778in}{0.000000in}}%
\pgfusepath{stroke,fill}%
}%
\begin{pgfscope}%
\pgfsys@transformshift{0.573273in}{1.759141in}%
\pgfsys@useobject{currentmarker}{}%
\end{pgfscope}%
\end{pgfscope}%
\begin{pgfscope}%
\pgfsetbuttcap%
\pgfsetroundjoin%
\definecolor{currentfill}{rgb}{0.000000,0.000000,0.000000}%
\pgfsetfillcolor{currentfill}%
\pgfsetlinewidth{0.602250pt}%
\definecolor{currentstroke}{rgb}{0.000000,0.000000,0.000000}%
\pgfsetstrokecolor{currentstroke}%
\pgfsetdash{}{0pt}%
\pgfsys@defobject{currentmarker}{\pgfqpoint{-0.027778in}{0.000000in}}{\pgfqpoint{0.000000in}{0.000000in}}{%
\pgfpathmoveto{\pgfqpoint{0.000000in}{0.000000in}}%
\pgfpathlineto{\pgfqpoint{-0.027778in}{0.000000in}}%
\pgfusepath{stroke,fill}%
}%
\begin{pgfscope}%
\pgfsys@transformshift{0.573273in}{1.874762in}%
\pgfsys@useobject{currentmarker}{}%
\end{pgfscope}%
\end{pgfscope}%
\begin{pgfscope}%
\pgfsetbuttcap%
\pgfsetroundjoin%
\definecolor{currentfill}{rgb}{0.000000,0.000000,0.000000}%
\pgfsetfillcolor{currentfill}%
\pgfsetlinewidth{0.602250pt}%
\definecolor{currentstroke}{rgb}{0.000000,0.000000,0.000000}%
\pgfsetstrokecolor{currentstroke}%
\pgfsetdash{}{0pt}%
\pgfsys@defobject{currentmarker}{\pgfqpoint{-0.027778in}{0.000000in}}{\pgfqpoint{0.000000in}{0.000000in}}{%
\pgfpathmoveto{\pgfqpoint{0.000000in}{0.000000in}}%
\pgfpathlineto{\pgfqpoint{-0.027778in}{0.000000in}}%
\pgfusepath{stroke,fill}%
}%
\begin{pgfscope}%
\pgfsys@transformshift{0.573273in}{1.974918in}%
\pgfsys@useobject{currentmarker}{}%
\end{pgfscope}%
\end{pgfscope}%
\begin{pgfscope}%
\pgfsetbuttcap%
\pgfsetroundjoin%
\definecolor{currentfill}{rgb}{0.000000,0.000000,0.000000}%
\pgfsetfillcolor{currentfill}%
\pgfsetlinewidth{0.602250pt}%
\definecolor{currentstroke}{rgb}{0.000000,0.000000,0.000000}%
\pgfsetstrokecolor{currentstroke}%
\pgfsetdash{}{0pt}%
\pgfsys@defobject{currentmarker}{\pgfqpoint{-0.027778in}{0.000000in}}{\pgfqpoint{0.000000in}{0.000000in}}{%
\pgfpathmoveto{\pgfqpoint{0.000000in}{0.000000in}}%
\pgfpathlineto{\pgfqpoint{-0.027778in}{0.000000in}}%
\pgfusepath{stroke,fill}%
}%
\begin{pgfscope}%
\pgfsys@transformshift{0.573273in}{2.063261in}%
\pgfsys@useobject{currentmarker}{}%
\end{pgfscope}%
\end{pgfscope}%
\begin{pgfscope}%
\pgftext[x=0.164322in,y=1.391720in,,bottom,rotate=90.000000]{\rmfamily\fontsize{10.000000}{12.000000}\selectfont \(\displaystyle E_k\)}%
\end{pgfscope}%
\begin{pgfscope}%
\pgfpathrectangle{\pgfqpoint{0.573273in}{0.415224in}}{\pgfqpoint{2.585875in}{1.952991in}} %
\pgfusepath{clip}%
\pgfsetrectcap%
\pgfsetroundjoin%
\pgfsetlinewidth{2.007500pt}%
\definecolor{currentstroke}{rgb}{0.900000,0.400000,0.400000}%
\pgfsetstrokecolor{currentstroke}%
\pgfsetdash{}{0pt}%
\pgfpathmoveto{\pgfqpoint{1.425830in}{1.792240in}}%
\pgfpathlineto{\pgfqpoint{1.472404in}{1.833046in}}%
\pgfpathlineto{\pgfqpoint{1.517666in}{1.869334in}}%
\pgfpathlineto{\pgfqpoint{1.561688in}{1.901443in}}%
\pgfpathlineto{\pgfqpoint{1.604536in}{1.929680in}}%
\pgfpathlineto{\pgfqpoint{1.646270in}{1.954323in}}%
\pgfpathlineto{\pgfqpoint{1.686948in}{1.975626in}}%
\pgfpathlineto{\pgfqpoint{1.726622in}{1.993820in}}%
\pgfpathlineto{\pgfqpoint{1.765339in}{2.009116in}}%
\pgfpathlineto{\pgfqpoint{1.803145in}{2.021708in}}%
\pgfpathlineto{\pgfqpoint{1.840083in}{2.031773in}}%
\pgfpathlineto{\pgfqpoint{1.876190in}{2.039476in}}%
\pgfpathlineto{\pgfqpoint{1.911503in}{2.044965in}}%
\pgfpathlineto{\pgfqpoint{1.946057in}{2.048380in}}%
\pgfpathlineto{\pgfqpoint{1.986564in}{2.049920in}}%
\pgfpathlineto{\pgfqpoint{2.026075in}{2.048861in}}%
\pgfpathlineto{\pgfqpoint{2.064637in}{2.045387in}}%
\pgfpathlineto{\pgfqpoint{2.102296in}{2.039669in}}%
\pgfpathlineto{\pgfqpoint{2.139092in}{2.031862in}}%
\pgfpathlineto{\pgfqpoint{2.175064in}{2.022109in}}%
\pgfpathlineto{\pgfqpoint{2.210249in}{2.010541in}}%
\pgfpathlineto{\pgfqpoint{2.250347in}{1.994912in}}%
\pgfpathlineto{\pgfqpoint{2.289468in}{1.977152in}}%
\pgfpathlineto{\pgfqpoint{2.327659in}{1.957421in}}%
\pgfpathlineto{\pgfqpoint{2.364964in}{1.935867in}}%
\pgfpathlineto{\pgfqpoint{2.406564in}{1.909172in}}%
\pgfpathlineto{\pgfqpoint{2.447113in}{1.880451in}}%
\pgfpathlineto{\pgfqpoint{2.486665in}{1.849871in}}%
\pgfpathlineto{\pgfqpoint{2.530026in}{1.813431in}}%
\pgfpathlineto{\pgfqpoint{2.572249in}{1.775021in}}%
\pgfpathlineto{\pgfqpoint{2.617896in}{1.730244in}}%
\pgfpathlineto{\pgfqpoint{2.662283in}{1.683467in}}%
\pgfpathlineto{\pgfqpoint{2.671015in}{1.673888in}}%
\pgfpathlineto{\pgfqpoint{2.671015in}{1.673888in}}%
\pgfusepath{stroke}%
\end{pgfscope}%
\begin{pgfscope}%
\pgfpathrectangle{\pgfqpoint{0.573273in}{0.415224in}}{\pgfqpoint{2.585875in}{1.952991in}} %
\pgfusepath{clip}%
\pgfsetrectcap%
\pgfsetroundjoin%
\pgfsetlinewidth{1.505625pt}%
\definecolor{currentstroke}{rgb}{0.000000,0.000000,0.000000}%
\pgfsetstrokecolor{currentstroke}%
\pgfsetdash{}{0pt}%
\pgfpathmoveto{\pgfqpoint{0.563273in}{0.525541in}}%
\pgfpathlineto{\pgfqpoint{0.589260in}{0.551975in}}%
\pgfpathlineto{\pgfqpoint{0.740947in}{0.772425in}}%
\pgfpathlineto{\pgfqpoint{0.866208in}{1.036683in}}%
\pgfpathlineto{\pgfqpoint{0.994812in}{1.246706in}}%
\pgfpathlineto{\pgfqpoint{1.224731in}{1.505767in}}%
\pgfpathlineto{\pgfqpoint{1.328425in}{1.688402in}}%
\pgfpathlineto{\pgfqpoint{1.425830in}{1.795736in}}%
\pgfpathlineto{\pgfqpoint{1.526567in}{1.889615in}}%
\pgfpathlineto{\pgfqpoint{1.604536in}{1.881164in}}%
\pgfpathlineto{\pgfqpoint{1.686948in}{1.973980in}}%
\pgfpathlineto{\pgfqpoint{1.765339in}{2.040415in}}%
\pgfpathlineto{\pgfqpoint{1.840083in}{2.033510in}}%
\pgfpathlineto{\pgfqpoint{1.911503in}{2.047257in}}%
\pgfpathlineto{\pgfqpoint{1.979883in}{2.048108in}}%
\pgfpathlineto{\pgfqpoint{2.045472in}{2.064094in}}%
\pgfpathlineto{\pgfqpoint{2.108487in}{2.059920in}}%
\pgfpathlineto{\pgfqpoint{2.169124in}{2.013297in}}%
\pgfpathlineto{\pgfqpoint{2.227556in}{2.010620in}}%
\pgfpathlineto{\pgfqpoint{2.283937in}{1.965483in}}%
\pgfpathlineto{\pgfqpoint{2.338407in}{1.935457in}}%
\pgfpathlineto{\pgfqpoint{2.391090in}{1.913395in}}%
\pgfpathlineto{\pgfqpoint{2.442100in}{1.856322in}}%
\pgfpathlineto{\pgfqpoint{2.491541in}{1.845259in}}%
\pgfpathlineto{\pgfqpoint{2.539506in}{1.815707in}}%
\pgfpathlineto{\pgfqpoint{2.586080in}{1.761637in}}%
\pgfpathlineto{\pgfqpoint{2.631342in}{1.736295in}}%
\pgfpathlineto{\pgfqpoint{2.718211in}{1.667508in}}%
\pgfpathlineto{\pgfqpoint{2.759946in}{1.616365in}}%
\pgfpathlineto{\pgfqpoint{2.800624in}{1.602630in}}%
\pgfpathlineto{\pgfqpoint{2.840297in}{1.582336in}}%
\pgfpathlineto{\pgfqpoint{2.879015in}{1.545028in}}%
\pgfpathlineto{\pgfqpoint{2.916821in}{1.517935in}}%
\pgfpathlineto{\pgfqpoint{2.953758in}{1.479009in}}%
\pgfpathlineto{\pgfqpoint{2.989865in}{1.468032in}}%
\pgfpathlineto{\pgfqpoint{3.025178in}{1.422445in}}%
\pgfpathlineto{\pgfqpoint{3.059732in}{1.396544in}}%
\pgfpathlineto{\pgfqpoint{3.093559in}{1.378074in}}%
\pgfpathlineto{\pgfqpoint{3.126688in}{1.354864in}}%
\pgfpathlineto{\pgfqpoint{3.159147in}{1.328697in}}%
\pgfpathlineto{\pgfqpoint{3.169147in}{1.315796in}}%
\pgfpathlineto{\pgfqpoint{3.169147in}{1.315796in}}%
\pgfusepath{stroke}%
\end{pgfscope}%
\begin{pgfscope}%
\pgfpathrectangle{\pgfqpoint{0.573273in}{0.415224in}}{\pgfqpoint{2.585875in}{1.952991in}} %
\pgfusepath{clip}%
\pgfsetbuttcap%
\pgfsetroundjoin%
\definecolor{currentfill}{rgb}{0.900000,0.400000,0.400000}%
\pgfsetfillcolor{currentfill}%
\pgfsetlinewidth{1.003750pt}%
\definecolor{currentstroke}{rgb}{0.900000,0.400000,0.400000}%
\pgfsetstrokecolor{currentstroke}%
\pgfsetdash{}{0pt}%
\pgfsys@defobject{currentmarker}{\pgfqpoint{-0.041667in}{-0.041667in}}{\pgfqpoint{0.041667in}{0.041667in}}{%
\pgfpathmoveto{\pgfqpoint{0.000000in}{-0.041667in}}%
\pgfpathcurveto{\pgfqpoint{0.011050in}{-0.041667in}}{\pgfqpoint{0.021649in}{-0.037276in}}{\pgfqpoint{0.029463in}{-0.029463in}}%
\pgfpathcurveto{\pgfqpoint{0.037276in}{-0.021649in}}{\pgfqpoint{0.041667in}{-0.011050in}}{\pgfqpoint{0.041667in}{0.000000in}}%
\pgfpathcurveto{\pgfqpoint{0.041667in}{0.011050in}}{\pgfqpoint{0.037276in}{0.021649in}}{\pgfqpoint{0.029463in}{0.029463in}}%
\pgfpathcurveto{\pgfqpoint{0.021649in}{0.037276in}}{\pgfqpoint{0.011050in}{0.041667in}}{\pgfqpoint{0.000000in}{0.041667in}}%
\pgfpathcurveto{\pgfqpoint{-0.011050in}{0.041667in}}{\pgfqpoint{-0.021649in}{0.037276in}}{\pgfqpoint{-0.029463in}{0.029463in}}%
\pgfpathcurveto{\pgfqpoint{-0.037276in}{0.021649in}}{\pgfqpoint{-0.041667in}{0.011050in}}{\pgfqpoint{-0.041667in}{0.000000in}}%
\pgfpathcurveto{\pgfqpoint{-0.041667in}{-0.011050in}}{\pgfqpoint{-0.037276in}{-0.021649in}}{\pgfqpoint{-0.029463in}{-0.029463in}}%
\pgfpathcurveto{\pgfqpoint{-0.021649in}{-0.037276in}}{\pgfqpoint{-0.011050in}{-0.041667in}}{\pgfqpoint{0.000000in}{-0.041667in}}%
\pgfpathclose%
\pgfusepath{stroke,fill}%
}%
\begin{pgfscope}%
\pgfsys@transformshift{1.425830in}{1.795736in}%
\pgfsys@useobject{currentmarker}{}%
\end{pgfscope}%
\end{pgfscope}%
\begin{pgfscope}%
\pgfpathrectangle{\pgfqpoint{0.573273in}{0.415224in}}{\pgfqpoint{2.585875in}{1.952991in}} %
\pgfusepath{clip}%
\pgfsetbuttcap%
\pgfsetroundjoin%
\definecolor{currentfill}{rgb}{0.900000,0.400000,0.400000}%
\pgfsetfillcolor{currentfill}%
\pgfsetlinewidth{1.003750pt}%
\definecolor{currentstroke}{rgb}{0.900000,0.400000,0.400000}%
\pgfsetstrokecolor{currentstroke}%
\pgfsetdash{}{0pt}%
\pgfsys@defobject{currentmarker}{\pgfqpoint{-0.041667in}{-0.041667in}}{\pgfqpoint{0.041667in}{0.041667in}}{%
\pgfpathmoveto{\pgfqpoint{0.000000in}{-0.041667in}}%
\pgfpathcurveto{\pgfqpoint{0.011050in}{-0.041667in}}{\pgfqpoint{0.021649in}{-0.037276in}}{\pgfqpoint{0.029463in}{-0.029463in}}%
\pgfpathcurveto{\pgfqpoint{0.037276in}{-0.021649in}}{\pgfqpoint{0.041667in}{-0.011050in}}{\pgfqpoint{0.041667in}{0.000000in}}%
\pgfpathcurveto{\pgfqpoint{0.041667in}{0.011050in}}{\pgfqpoint{0.037276in}{0.021649in}}{\pgfqpoint{0.029463in}{0.029463in}}%
\pgfpathcurveto{\pgfqpoint{0.021649in}{0.037276in}}{\pgfqpoint{0.011050in}{0.041667in}}{\pgfqpoint{0.000000in}{0.041667in}}%
\pgfpathcurveto{\pgfqpoint{-0.011050in}{0.041667in}}{\pgfqpoint{-0.021649in}{0.037276in}}{\pgfqpoint{-0.029463in}{0.029463in}}%
\pgfpathcurveto{\pgfqpoint{-0.037276in}{0.021649in}}{\pgfqpoint{-0.041667in}{0.011050in}}{\pgfqpoint{-0.041667in}{0.000000in}}%
\pgfpathcurveto{\pgfqpoint{-0.041667in}{-0.011050in}}{\pgfqpoint{-0.037276in}{-0.021649in}}{\pgfqpoint{-0.029463in}{-0.029463in}}%
\pgfpathcurveto{\pgfqpoint{-0.021649in}{-0.037276in}}{\pgfqpoint{-0.011050in}{-0.041667in}}{\pgfqpoint{0.000000in}{-0.041667in}}%
\pgfpathclose%
\pgfusepath{stroke,fill}%
}%
\begin{pgfscope}%
\pgfsys@transformshift{2.675363in}{1.702015in}%
\pgfsys@useobject{currentmarker}{}%
\end{pgfscope}%
\end{pgfscope}%
\begin{pgfscope}%
\pgfsetrectcap%
\pgfsetmiterjoin%
\pgfsetlinewidth{0.803000pt}%
\definecolor{currentstroke}{rgb}{0.000000,0.000000,0.000000}%
\pgfsetstrokecolor{currentstroke}%
\pgfsetdash{}{0pt}%
\pgfpathmoveto{\pgfqpoint{0.573273in}{0.415224in}}%
\pgfpathlineto{\pgfqpoint{0.573273in}{2.368215in}}%
\pgfusepath{stroke}%
\end{pgfscope}%
\begin{pgfscope}%
\pgfsetrectcap%
\pgfsetmiterjoin%
\pgfsetlinewidth{0.803000pt}%
\definecolor{currentstroke}{rgb}{0.000000,0.000000,0.000000}%
\pgfsetstrokecolor{currentstroke}%
\pgfsetdash{}{0pt}%
\pgfpathmoveto{\pgfqpoint{3.159147in}{0.415224in}}%
\pgfpathlineto{\pgfqpoint{3.159147in}{2.368215in}}%
\pgfusepath{stroke}%
\end{pgfscope}%
\begin{pgfscope}%
\pgfsetrectcap%
\pgfsetmiterjoin%
\pgfsetlinewidth{0.803000pt}%
\definecolor{currentstroke}{rgb}{0.000000,0.000000,0.000000}%
\pgfsetstrokecolor{currentstroke}%
\pgfsetdash{}{0pt}%
\pgfpathmoveto{\pgfqpoint{0.573273in}{0.415224in}}%
\pgfpathlineto{\pgfqpoint{3.159147in}{0.415224in}}%
\pgfusepath{stroke}%
\end{pgfscope}%
\begin{pgfscope}%
\pgfsetrectcap%
\pgfsetmiterjoin%
\pgfsetlinewidth{0.803000pt}%
\definecolor{currentstroke}{rgb}{0.000000,0.000000,0.000000}%
\pgfsetstrokecolor{currentstroke}%
\pgfsetdash{}{0pt}%
\pgfpathmoveto{\pgfqpoint{0.573273in}{2.368215in}}%
\pgfpathlineto{\pgfqpoint{3.159147in}{2.368215in}}%
\pgfusepath{stroke}%
\end{pgfscope}%
\begin{pgfscope}%
\pgfsetrectcap%
\pgfsetroundjoin%
\pgfsetlinewidth{2.007500pt}%
\definecolor{currentstroke}{rgb}{0.900000,0.400000,0.400000}%
\pgfsetstrokecolor{currentstroke}%
\pgfsetdash{}{0pt}%
\pgfpathmoveto{\pgfqpoint{1.308708in}{0.730646in}}%
\pgfpathlineto{\pgfqpoint{1.530930in}{0.730646in}}%
\pgfusepath{stroke}%
\end{pgfscope}%
\begin{pgfscope}%
\pgftext[x=1.619819in,y=0.691757in,left,base]{\rmfamily\fontsize{8.000000}{9.600000}\selectfont fitted parabola}%
\end{pgfscope}%
\begin{pgfscope}%
\pgfsetrectcap%
\pgfsetroundjoin%
\pgfsetlinewidth{1.505625pt}%
\definecolor{currentstroke}{rgb}{0.000000,0.000000,0.000000}%
\pgfsetstrokecolor{currentstroke}%
\pgfsetdash{}{0pt}%
\pgfpathmoveto{\pgfqpoint{1.308708in}{0.575712in}}%
\pgfpathlineto{\pgfqpoint{1.530930in}{0.575712in}}%
\pgfusepath{stroke}%
\end{pgfscope}%
\begin{pgfscope}%
\pgftext[x=1.619819in,y=0.536824in,left,base]{\rmfamily\fontsize{8.000000}{9.600000}\selectfont simulation data}%
\end{pgfscope}%
\end{pgfpicture}%
\makeatother%
\endgroup%